\documentclass[12pt, letterpaper]{article}
 
\usepackage{graphicx}
\usepackage{enumerate}
\usepackage[authoryear]{natbib}
\usepackage{url}  
\usepackage[english]{babel}
 \usepackage{color}
\usepackage{multirow}
\usepackage{lscape}
\usepackage{csquotes}
\usepackage{verbatim}
 \usepackage{prodint}
\usepackage[english]{babel}
\usepackage[utf8]{inputenc}
\usepackage{amsmath, amsthm, amssymb,amsfonts,bm}
\usepackage{tikz}
\usepackage{courier}
\usetikzlibrary{positioning,shapes.geometric,graphs, arrows.meta}
\usetikzlibrary[graphs]
\usepackage{color}
\usepackage{verbatim}
\usepackage{hyperref}
\usepackage{float}
\usepackage{comment}
\usepackage{mathrsfs}
\usepackage{multicol}
\usepackage{bbm}
\usepackage{lipsum}
  \usepackage{color}
\usepackage{graphicx,subcaption}  
\usepackage{float}
\usepackage{comment}
\usepackage{mathrsfs}
 \usepackage{color}
 
     \usepackage{titlesec}
 \usepackage{setspace}
\RequirePackage{amsthm,amsmath,amsfonts,amssymb}

\RequirePackage[authoryear]{natbib}
\usepackage[authoryear]{natbib}

\usepackage{geometry}
\begin{document}
 
\title{Estimation of the Marginal Effect of Antidepressants \\ on Body Mass Index under Confounding and  Endogenous \\ Covariate-Driven Monitoring Times}
 \author{Janie Coulombe, Erica E.M. Moodie, Robert W. Platt, \\ and Christel Renoux}
  \date{}
  \maketitle

\begin{abstract}
 
In studying the marginal effect of antidepressants on body mass index using electronic health records data, we face several challenges. Patients' characteristics can affect the exposure (confounding) as well as the timing of routine visits (measurement process), and those characteristics may be altered following a visit which can create dependencies between the monitoring and body mass index when viewed as a stochastic or random processes in time. This may result in a form of selection bias that distorts the estimation of the marginal effect of the antidepressant.
Inverse intensity of visit weights have been proposed to adjust for these imbalances, however no approaches have addressed complex settings where the covariate and the monitoring processes affect each other in time so as to induce endogeneity, a situation likely to occur in electronic health records. We review how selection bias due to outcome-dependent follow-up times may arise and propose a new cumulated weight that models a complete monitoring path so as to address the above-mentioned challenges and produce a reliable estimate of the impact of antidepressants on body mass index. More specifically, we do so using data from the Clinical Practice Research Datalink in the United Kingdom, comparing the marginal effect of two commonly used antidepressants, citalopram and fluoxetine, on body mass index. The results are compared to those obtained with simpler methods that do not account for the extent of the dependence due to an endogenous covariate process. 
\end{abstract}

   \section{Introduction} 
 
Citalopram and fluoxetine are two selective serotonin reuptake inhibitors (SSRIs) commonly prescribed as a first-line treatment in patients with depression. Weight gain is a side effect associated with the use of antidepressants (including SSRIs) which may lead to treatment discontinuation or non-adherence (see e.g. De las Cuevas, Pen\~ate and Sanz, \citeyear{de2014risk}). The literature is undecided on whether there is a differential effect of these two drugs on weight (Sussman and Ginsberg, \citeyear{sussman1998rethinking}; Serretti et al., \citeyear{serretti2010antidepressants}). No previous studies using electronic health records (EHR) data have compared their effect on weight or body mass index (BMI) after adjusting for potential confounders and selection bias due to outcome-dependent follow-up time. However, EHR and administrative databases do not collect data for research purposes, and patient BMI is not recorded at every follow-up visit. Its recording may be associated with patient characteristics.

When making inference on the causal effect of an exposure on a longitudinal outcome that is measured repeatedly over time using observational data, the analyst only observes data according to a measurement frequency that is driven by the characteristics of the individuals under observation. For instance, the smoking status of a patient, their comorbidities and general health state, their prescription drugs, or other characteristics can affect the  frequency with which they visit their physician, and these same characteristics can change following a routine visit, as the physician may recommend lifestyle changes or the mere fact of having a visit may increase a patient's awareness of their habits. The imbalances related to the visit process can lead to biased estimates of the exposure effect in situations where the monitoring indicator is a collider and that conditioning on it opens a path from the exposure to the longitudinal outcome. This phenomenon is sometimes referred to as outcome-dependent follow-up times (Lipsitz, Fitzmaurice, Ibrahim, et al., \citeyear{lipsitz2002parameter}).

Several methods have been proposed to handle imbalances due to outcome-dependent follow-up, with most discussed outside of a causal framework. Some authors proposed joint modelling of the monitoring and the outcomes processes via random effects or shared parameters (see e.g. Liang et al., \citeyear{liang2009joint}). Pullenayegum (\citeyear{pullenayegum2016multiple}) discussed outputation, an approach that thins the visit process by repeatedly discarding selected observations, to make the monitoring process independent of covariates. Pullenayegum and Lim (\citeyear{pullenayegum2016longitudinal}) also presented an insightful review on methods for irregular visits, including inverse intensity of visit weights (IIVW). The proposed IIVW of Lin, Scharfstein and Rosenheck (\citeyear{lin2004analysis}) apply to settings in which the monitoring rate is modelled as a function of exogenous covariates. It is particularly convenient to use inverse weighting in the context of causal inference on the marginal effect of exposure, as the weights can easily be incorporated into  estimating equations to obtain an estimate of treatment effect, and they can be function of mediators of the exposure-outcome effect that affect monitoring.

Monitoring times and the outcome process are both stochastic processes which occur in continuous time. In the estimation of the marginal effect of antidepressants on BMI, a common set of characteristics affects the monitoring times and the BMI values, and the dependence between the monitoring and the BMI processes may be complex. Yet, to date, no weighting strategy has been proposed to address the difficult problem of an \textit{endogenous} covariate process interacting with the monitoring process and the outcome process in continuous time in a setting where monitoring times are informative (note, a covariate process is said to be endogenous if the covariate is affected by the fact of there being a visit, which can create long-term dependencies between the monitoring and the outcome processes). Beyond the research question  that motivates this work, several other examples of such processes exist, including the setting in which monitoring times allow a physician to take a treatment decision and prescribe a new treatment which may affect a patient's future outcome and monitoring time, or that where the advice of the physician may affect a patient's future habits (such as smoking). In all each of these examples, the gap time between visits may also act as an endogenous process that is modified by a new visit (and returns to zero when there is such visit). If the gap time is affected by variables that make the monitoring and the outcome processes dependent in time, this too may induce biasing dependencies.

In this work, we focus on the estimation of the (causal) marginal effect of two antidepressants, citalopram and fluoxetine, on BMI  which is measured repeatedly over time. We wish to address the possibility that the relationship of interest is distorted by imbalances due to the monitoring process and its relation with the BMI process, as well as confounding, which two features were never considered simultaneously in the assessment of the marginal effect of antidepressants on weight or BMI. Our methodology generalizes easily to the estimation of the causal marginal effect of a binary intervention, on a longitudinal continuous outcome, in observational studies. We provide a thorough description of the bias due to covariate-driven monitoring times in longitudinal settings, with a demonstration that relies on causal diagrams. In addition, we propose the first weighting method for addressing situations with longitudinal collider stratification bias (Greenland, \citeyear{greenland2003quantifying}) that is due to an endogenous covariate process affecting the monitoring times. Another key contribution of our method, as opposed to those previously proposed, is that it loosens the restrictions on when the covariate process (affecting and being affected by monitoring process) has to be observed; standard inverse intensity of visit weights rely on the assumption that the monitoring model's covariates are assessed in continuous time while our methodology uses modelling of the gap times as functions of the last covariates observed, throughout patients' follow-up. This represents an important weakening of the assumptions needed for consistent estimation.

This paper is structured as follows: in Section 2, we present the notation and assumptions, some background and the proposed extension. In Section 3, we present details on the simulation studies that were conducted to assess the performance of the proposed methods along with their results. In Section 4, we present the analysis of the data from the Clinical Practice Research Datalink (CPRD) (Herrett et al., \citeyear{herrett2015data}), where we compare different estimators for the marginal effects of citalopram and fluoxetine on BMI using the novel weighting method. A discussion follows in Section 5.

\section{Methods}
\subsection{Notation and Causal Assumptions}\label{as1}

Let $i$ denote the index of an individual, $i=1,...,n$, and $t$ denote the time, which is continuous, with $t \in [0, \tau]$ and $\tau$  the maximum follow-up time in the cohort under study. We are interested in estimating the causal marginal effect of a binary point intervention $I_i(0)$ (i.e., a choice between two antidepressants) on the continuous, longitudinal outcome $Y_i(t)$ (BMI). Our interest is in a time-fixed (``point'') intervention for two antidepressant drugs, however extensions to the weighting approach for the time-varying intervention case are straightforward. Bold is used to refer to vectors and matrices.

We use the Neyman-Rubin potential outcome framework (Neyman, \citeyear{neyman1923application}; Rubin, \citeyear{rubin1974estimating}) to express the estimand of interest, which is the causal contrast $\mathbb{E}\left[Y_{i1}(t) - Y_{i0}(t)\right]$ where $Y_{i1}(t)$ corresponds to the outcome that would have been observed at time $t$, had individual $i$ received intervention $I_i(0)=1$, and $Y_{i0}(t)$, had the individual received intervention $I_i(0)=0$. In the analysis of the CPRD data, this contrast corresponds to the average difference in BMI had everyone been treated with citalopram, versus had everyone been treated with fluoxetine. Our interest is in the time-invariant effect of citalopram vs fluoxetine (i.e.~we assume that the impact is not modified by time: $\mathbb{E}\left[Y_{i1}(t) - Y_{i0}(t)\right] = \theta$).

We are interested in addressing two important sources of bias that may distort estimators of our parameter of interest. First, we consider confounding bias, where confounders, denoted by $\mathbf{K(t)}= \mathbf{K(0)}$, are covariates that affect both the intervention and the outcome $Y_i(t)$ under study. As the intervention is time-fixed, confounders are measured at baseline, prior to receiving the intervention. Second, we consider selection bias caused by covariate-informed monitoring times (also referred to as \textit{visit times} throughout). In this work, individuals are allowed to have completely different sets of monitoring times (and thus, a unique pattern of visits). In practice, we choose a certain coarsening of time (e.g., daily) over which we observe monitoring times. 
To distinguish these two characterizations of time, we will assume, when needed, that the coarsening is daily and will denote discrete times by $t=0$ (baseline), $1, 2, 3,..., \tau$. For continuous time ($t \in [0, \tau]$), we will use the notation $t-$ to refer to the moment immediately before time $t$.

We suppose that monitoring times coincide with the observation of the longitudinal continuous outcome $Y_i(t)$. This means that we consider a monitoring time to be any time, and only those times, when BMI is assessed. We denote by $N_i(t)$ a counting process for monitoring times in individual $i$, which counts the number of previous visits they had by time $t$. A monitoring indicator at time $t$ is denoted by $dN_i(t)$ in individual $i$. We also introduce the notation $l_i(t)=t-B_i(t)$, with $l_i(t)$ the previous (most recent) visit time of individual $i$ at time $t$, where $B_i(t)$ is a time-dependent gap time which gives, at time $t$, the delay since the last visit in individual $i$. For each individual, let $C_i$ denote the censoring time, that is, the time until which we can potentially observe an individual, their covariate and their outcome processes. Let $\xi_i(t)= \mathbb{I}(C_i\ge t)$ be the indicator of individual $i$ still being in the study at time $t$. We assume throughout that censoring is uninformative except through the monitoring process. That is, we assume that we capture in modelling the monitoring process any (possibly biasing) imbalances in censoring times across the treatment groups.
Denote by $\mathbf{\mathcal{H}^o_i(t-)}$  the \textit{observed} history of covariates (personal characteristics, monitoring indicators, outcome values; these will be discussed in depth in Section \ref{analysis}, for the analysis of CPRD data) by time $t-$ in individual $i$. Since the intervention is given at baseline, $\mathbf{\mathcal{H}^o_i(t-)}$ may contain mediators of the relationship between the intervention and the longitudinal outcome $Y_i(t)$. We acknowledge that  monitoring indicators $dN_i(t), t \in [0, \tau]$ can be colliders and block the path between the intervention and the outcome of interest throughout follow-up time. Conditioning on these colliders by using only the observed data can unblock the path between intervention and outcome and risks biasing the estimator of the marginal effect of intervention of interest.

We assume:
\begin{align}
\tag{I1} I_i(0) \perp \left\{ Y_{i0}(t), Y_{i1}(t)\right\} | &   \mathbf{K_i(0)}  \label{v0} \\
\tag{I2} I_i(0) \not\perp \left\{ Y_{i0}(t), Y_{i1}(t)\right\} | &  dN_i(t), \mathbf{K_i(0)}  \label{v1} \\
\tag{I3} I_i(0) \perp \left\{ Y_{i0}(t), Y_{i1}(t)\right\} |  &\mathbf{\mathcal{H}^o_i(t-)} , dN_i(t), \mathbf{K_i(0)} \label{v2}\\
\tag{P1} 0<P(dN_i(t)=1 |\mathbf{K_i(0)}, &\mathbf{\mathcal{H}^o_i(t-)}), P(dN_i(t)=0 |\mathbf{K_i(0)}, \mathbf{\mathcal{H}^o_i(t-)}) <1 \label{p4}  \\
\tag{P2} 0<P(I_i(0)=1 | \mathbf{K_i(0)}),& P(I_i(0)=0 | \mathbf{K_i(0)})<1, \label{p3}
\end{align}
such that $\mathbf{\mathcal{H}^o_i(t-)}$ and $\mathbf{K_i(0)}$ are sufficient sets to break the dependency between the potential outcome and the intervention, even when conditioning on the potential collider $dN_i(t)$. We further assume that we have positivity for intervention and monitoring (assumptions \ref{p4} and \ref{p3}), where positivity for monitoring essentially means that there is at least a non-zero probability for any patient to have a visit on any given day (or week, depending on time granularity) while there is no day when it is 100\% certain that a visit will occur. We also assume the stable unit treatment value assumption (SUTVA), a condition that encompasses a well-defined exposure or intervention, as well as no interference between individuals' effects.

Suppose that both the mean outcome process and the monitoring process at time $t$ may depend on the confounders $\mathbf{K(0)}$, the intervention $\mathbf{I(0)}$, as well as a longitudinal, possibly vector-valued covariate process $\mathbf{Z_i(l_i(t))} \subset \mathbf{\mathcal{H}^o_i(t-)}$ that may be affected by monitoring times in the past.  
Like the outcome, we assume that this additional covariate process may only be assessed at monitoring times  
(such that $Z_i(t)=Z_i(l_i(t)) \hspace{0.1cm}\forall t$), a not unrealistic scenario.  
We further assume the following conditional outcome mean model
\begin{align}
\tag{CO} \mathbb{E}\left[ Y_i(t) | \mathbf{K_i(0)}, \mathbf{Z_i(l_i(t))} , I_i(0)\right] = \alpha_0(t) + \bm{\beta_K} \mathbf{K_i(0)} + \bm{\beta_Z} \mathbf{Z_i(l_i(t))} + \beta_I I_i(0),\label{o2}
\end{align}
with $\alpha_0(t)$ a flexible intercept function. The variables $\mathbf{Z(l_i(t))}$ may include mediators of the relationship between $I_i(0)$ and $Y_i(t)$ (in the context of our analysis, for example, a variable like having a diagnosis of diabetes is affected by antidepressant drugs and may itself affect individuals' weight). We aim to estimate the total effect of the intervention, without distortion by confounding or blocking paths which act through mediators. We henceforth use weights to create a pseudo-population in which there is no imbalance between confounder variables across intervention groups, and in which the monitoring and the outcome processes are independent. Using data from that pseudo-population, one can further use the marginal outcome mean model that follows
\begin{align}
\tag{MO} \mathbb{E}\left[ Y_i(t) | I_i(0) \right] = \alpha(t) +  \beta I_i(0);\label{o3}
\end{align}
$\beta$ in (\ref{o3}) represents the average intervention effect of $\mathbf{I(0})$ for which we seek an estimate. Denote by $\alpha(t)$ the intercept in the model, which may vary with time. We will also consider scenarios in which the intercept varies by individual ($\alpha_i(t)$). 

In settings with outcome-dependent follow-up, Lin et al.~(2004) have proposed the following continuous-time estimating equation to estimate the effect $\bm{\beta_0}$ of some time-fixed covariates $\mathbf{X}$ on an outcome $\mathbf{Y(t)}$ (which is not necessarily continuous):
\begin{align*}
\mathbb{E}\left[ \int_0^{\tau} \left\{ \mathbf{Y(t)} -\bm{\mu}(\bm{t}, \mathbf{X}; \bm{\beta_0})\right\} \frac{\mathbf{c}(\bm{t,} \mathbf{X}; \bm{\beta_0)}}{ \bm{\lambda}(\bm{t} |\mathbf{\mathcal{H}^o(t-)} ) } \mathbf{dN(t)}\right] =\mathbf{0},
\end{align*}
in which $\mathbf{c}(\bm{t}, \mathbf{X}; \bm{\beta_0})$ is an arbitrary weight, $\bm{\mu}(\bm{t}, \mathbf{X}; \bm{\beta_0})$ a mean function, and $\bm{\lambda}(\bm{t} |\mathbf{\mathcal{H}^o(t-)} )$, an IIVW which is a  function of the history of observed variables.

Similarly, Coulombe et al. (\citeyear{coulombe}) proposed two flexible estimators for the marginal effect of the intervention; these accounted for confounding along with informative monitoring times and were tested in settings where the intervention can vary in time.  In the context that interests us, where we aim to estimate $\beta$ in the marginal outcome mean model (\ref{o3}), the corresponding design matrix ($\mathbf{X(t)}$) is composed of the intervention $\mathbf{I(0)}$ and may also contain functions of time such as a cubic spline basis to model the intercept function $\alpha(t)$. However, the authors did not  
consider individualized intercepts that can vary in time, a setting which raises additional challenges, and no previous author has considered endogeneity of the covariate process, which can create long-term dependence in the monitoring path that goes beyond the current covariates. Further, in both Bůžková and Lumley (\citeyear{buuvzkova2009semiparametric}) and Coulombe et al. (\citeyear{coulombe}), it is assumed that variables that affect monitoring time are observed at all times, a frequently unrealistic assumption. In the estimation of the marginal effect of antidepressants on BMI, in particular, considering these features will allow more flexibility. It will further allow us to postulate weaker assumptions on the monitoring process. In the CPRD data, patients' characteristics that lead them to visiting (or not) their physician are unlikely to be updated in continuous time; rather, their measurement mostly coincides with that of the outcome, BMI.

For the inverse weight (and intensity function) $\bm{\lambda}(\bm{t} |\mathbf{\mathcal{H}^o(t-)} )$, different modelling strategies have been considered by previous authors, such as different time scales or conditioning on different sets of covariates (Bůžková and Lumley, \citeyear{buuvzkova2009semiparametric}; Zhu et al., \citeyear{zhu2017estimation}; Coulombe et al., \citeyear{coulombe}).  The intensity for a counting process is defined by the instantaneous rate, which is given by $\lambda_i(t |\mathbf{\mathcal{H}^o_i(t-)} ) =  \lim_{dt \rightarrow 0} P(N_i(t+ dt) - N_i(t) =1  | \mathbf{\mathcal{H}^o_i(t-)}) / dt$. The rate is preferred to a discrete probability model, as the monitoring ``events'' can occur at any time on a continuous time scale. In practice, and assuming that as $dt$ gets closer to 0, the time units are so small that only one jump can occur per time unit $dt$, one can view this as a Bernoulli experiment over each small time unit $dt$ with a certain probability of visit.  

By definition, a conditional intensity model uniquely defines the counting process and its dependency on the past (Lindsey, \citeyear{lindsey2004statistical}; Cook and Lawless, \citeyear{cook2007statistical}), including previous monitoring times.  
The model may be affected by whether or not the covariates affecting the intensity function are endogenous. If they are exogenous, and if monitoring at time $t$ does not depend on previous monitoring times, a marginal approach where the marginal effect of covariates is estimated will suffice (e.g.,~as proposed in Lin et al., \citeyear{lin2000semiparametric}). If the covariate process is endogenous and if visit at time $t$ depends on both the covariate process interacting with it and previous monitoring times, the conditional intensity function may have to account for complex functions of the past. In particular, links between the covariate, the monitoring and the outcome processes can exist due to the endogeneity.

\subsection{Visit Process Scenarios and their Data Generating Mechanisms}

We now propose different general scenarios for the monitoring times process, and describe the associated data generating mechanisms (DGMs).  
The first two DGMs refer to scenarios often encountered (and postulated) in the literature. The third and fourth DGMs correspond to situations where monitoring times affect the endogenous covariate process, such as we postulate in our analysis of the CPRD data. They are used to demonstrate the potential selection bias and the proposed methodology. 

For each DGM, we review how the selection bias due outcome-dependent follow-up times arises, and which IIVW can be incorporated in the estimating equations for the marginal effect of intervention to make correct inferences. In each diagram, we depict all time points over which a visit can occur. We also assume that bias due to confounder variables is appropriately accounted for via classic adjustment methods such as the inverse probability of treatment (intervention) weight (Rosenbaum and Rubin, \citeyear{rosenbaum1983central}).

The first DGM we consider is depicted in Figure \ref{fig1} and is reminiscent of the causal diagram considered in Coulombe et al. (\citeyear{coulombe}) in which the marginal intensity of a visit at time $t$ can be modelled using exogenous covariates measured at time $t$. Suppose that we aim to estimate the causal effect of the baseline intervention $I(0)$ on the longitudinal outcome $Y(t)$. In that DGM, at each time $t$, $Z(t)$ acts as a mediator of the relationship of interest. The selection bias due to outcome-dependent follow-up times comes from those mediators, as conditioning on $dN(t)$, $t\ge0$ unblocks the path going through colliders at each node $dN(t)$ ($t \in 0, 1$), which opens the path $I(0) - dN(t) - Z(t) \rightarrow Y(t)$ $\forall t$.

In that setting, to address the selection bias due to conditioning on the observed data, one can fit a proportional rate model for the rate of visit while conditioning on both $I(0)$ and $Z(t)$ at each time $t$ as we have that $dN(t) \perp Y(t) | I(0), Z(t), \mathbf{K(0)}$. Further, there is no dependency structure across monitoring times that must be considered. We assume 
\begin{align}
 \mathbb{E}\left[ dN_i(t)| I_i(0), Z_i(t) \right] =\xi_i(t) \exp \left( \gamma_I I_i(0) + \gamma_Z Z(t)\right) \lambda_0(t) dt, \label{vp1}
\end{align}
a proportional rate model where the effect of covariates is captured via the $\bm{\gamma}=\left\{\gamma_I, \gamma_Z \right\}$ parameters, and the effect of time via $\lambda_0(t)$. For the estimation of the $\bm{\gamma}$ parameters, one can use the Andersen and Gill model (\citeyear{andersen1982cox}), an extension of the Cox proportional hazards model for recurrent events. The baseline rate $\lambda_0(t)$ need not be estimated if the time scale is time since cohort entry, as the term will cancel out across individuals (Bůžková and Lumley, \citeyear{buuvzkova2009semiparametric}). If the time scale is otherwise (e.g., if it is the gap time, $B_i(t))$, then the weight should incorporate a baseline function of that alternate time scale (Zhu et al., \citeyear{zhu2017estimation}); Breslow-type estimators can be used to estimate the baseline rate (Cox, \citeyear{cox1972regression}). For fitting the model in (\ref{vp1}), covariates $\left\{ I_i(0), Z_i(t)\right\}$ must be measured (recorded) at all times during each patient's follow-up, which is not necessarily straightforward in observational longitudinal studies.

The second DGM we consider is presented in Figure \ref{fig2} and is similar to the DGM considered in Zhu et al. (\citeyear{zhu2017estimation}). As compared to the causal diagram in Figures \ref{fig1},  covariates measured \textit{before time $t$} are now assumed to affect both the monitoring and the outcome at time $t$, and to mediate the effect of interest. In this second DGM scenario, conditioning on colliders $dN(\cdot)$ by analyzing available data unblocks a path from the intervention and the outcome, via e.g.~$I(0) - dN(1) -  Z(0) \rightarrow Y(1) $, or, similarly, via $I(0) - dN(1) -  Z(0) \rightarrow Y(2) $. A proportional rate model can be used for blocking these biasing paths from the intervention to the outcome. We must now condition on covariates measured or updated in the past to adjust for selection bias, and must assume that no covariate measured after that point in the past affects both the next outcome and monitoring indicator. Suppose that we denote by $Z(t-)$ the last covariate value of $Z(\cdot)$ that affects, at time $t$, the monitoring indicator and the outcome; then the following model for monitoring will appropriately address the covariate-dependent monitoring times:
\begin{align}
 \mathbb{E}\left[ dN_i(t)| I_i(0), Z_i(t-) \right] =& \xi_i(t) \exp \left( \gamma_I I_i(0) + \gamma_Z Z_i(t-)\right) \lambda_0(t) dt. \label{vp2}
\end{align}
 \begin{minipage}[t]{0.5\textwidth}
\centering
\begin{tikzpicture}[scale=0.50][%
->,
shorten >=2pt,
>=stealth,
node distance=1cm,
pil/.style={
->,
thick,
shorten =2pt,}
]


\node(1) at (-4,0){\textcolor{black}{$\mathbf{K(0)}$}};


\node (2) at (-1,2) {$ I(0) $};

 \node(3) at (-1,-7) {$Y(0)$};
 
 \node(3c) at (5.5,-7) {$Y(1)$};

\node(4) at (1, -3) {$Z(0)$};
 
\node(4c) at (5, -2.8) {$Z(1)$};

\node(5) at (1.2,-8) {dN(0)};
 
\node(5c) at (7.3,-6) {dN(1)};

\draw[-{Latex[length=3mm]}] (1) to (2);
\draw[-{Latex[length=3mm]}] (1) to (3);

\draw[-{Latex[length=3mm]}] (1) to (3c);

\draw[-{Latex[length=3mm]}] (2) to (3);
 
\draw[-{Latex[length=3mm]}] (2) to (3c);

\draw[-{Latex[length=3mm]}] (2) to (4);
 \draw[-{Latex[length=3mm]}] (4) to (3);

\draw[-{Latex[length=3mm]}] (2) to (4c);
 \draw[-{Latex[length=3mm]}] (4c) to (3c);

\draw[-{Latex[length=3mm]}] (2) to (5);
 
\draw[-{Latex[length=3mm]}] (2) to (5c);
\draw[-{Latex[length=3mm]}] (4) to (5);
 
\draw[-{Latex[length=3mm]}] (4c) to (5c);

\end{tikzpicture} 
\captionsetup{font=footnotesize}

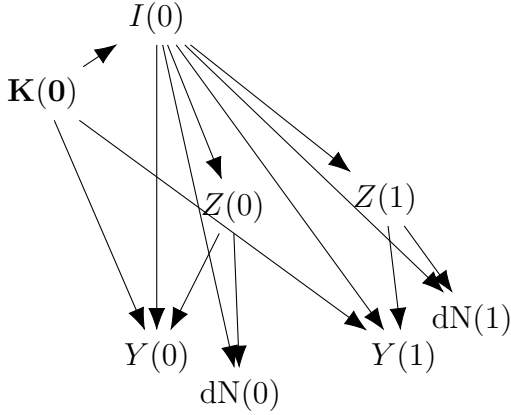
\captionof{figure}{Causal diagram for the first DGM (patient index $i$ removed) $I(0)$ is an intervention of interest whose marginal effect on a longitudinal outcome, $Y(t)$ -- assumed to be time-invariant -- is of interest. $\mathbf{K(0)}$ represent confounding variables, $Z(t)$ are mediators, and $dN(t)$ indicates the monitoring process through which the outcome is observed.}\label{fig1}
\end{minipage}\hspace{.6cm}
\begin{minipage}[t]{0.5\textwidth}
 \centering
\begin{tikzpicture}[scale=0.45][%
->,
shorten >=2pt,
>=stealth,
node distance=1cm,
pil/.style={
->,
thick,
shorten =2pt,}
]

\node (1) at (-1.4,-2.5) {$ I(0) $};

 \node(3b) at (5,0) {$Y(1)$};
\node(3c) at (9.5,0.3) {$Y(2)$};
\node(3d) at (14,-0) {$Y(3)$};
\node(5) at (2.5,3.4){\textcolor{black}{$\mathbf{K(0)}$}};

\draw[-{Latex[length=3mm]}] (1) to  (3b);
\draw[-{Latex[length=3mm]}] (1) to  (3c);
\draw[-{Latex[length=3mm]}](5) to  (1);

 \draw[-{Latex[length=3mm]}](5) to (3b);
\draw[-{Latex[length=3mm]}](5) to (3c);

\node(30) at (-1.5, -6.3) {$Z(0)$};

\node(307) at (9, -4) {$Z(2)$};

\draw[-{Latex[length=3mm]}](1) to (30);

 \draw[-{Latex[length=3mm]}] (307) to (3d);

\node(11) at (4,-9) {\textcolor{black}{$dN(1)$}};
 \node(12) at (8.5,-8) {\textcolor{black}{$dN(2)$}};
   \node(13) at (13,-8.3) {\textcolor{black}{$dN(3)$}};
 \draw[-{Latex[length=3mm]}] (307) to (13);
\draw[-{Latex[length=3mm]}](5) to (3d);

 \draw[-{Latex[length=3mm]}] (1) to (307);
\draw[-{Latex[length=3mm]}](1) to (11);
\draw[-{Latex[length=3mm]}](1) to (12);
\draw[-{Latex[length=3mm]} ](30) to (11);
\draw[-{Latex[length=3mm]}](1) to (3d);

\draw[-{Latex[length=3mm]}](30) to (3b);

 \draw[-{Latex[length=3mm]}](30) to (3c);
 \draw[-{Latex[length=3mm]}](30) to (12);

 \draw[-{Latex[length=3mm]}] (1) to (13);

\end{tikzpicture} 
\captionsetup{font=footnotesize}

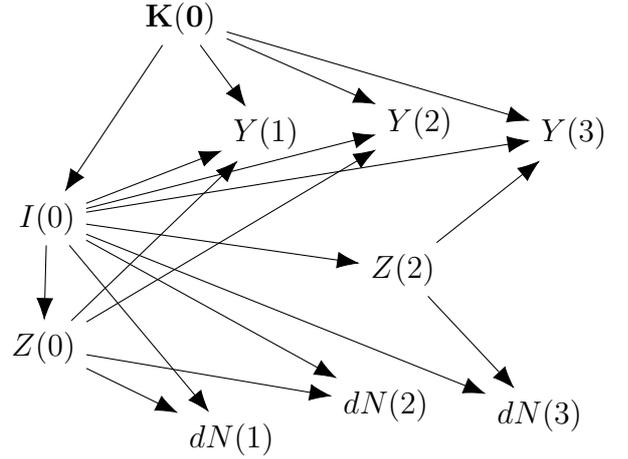
\captionof{figure}{Causal diagram for the second DGM (patient index $i$ removed) $I(0)$ is an intervention of interest whose marginal effect on a longitudinal outcome, $Y(t)$ -- assumed to be time-invariant -- is of interest. $\mathbf{K(0)}$ represent confounding variables, $Z(t)$ are mediators, and $dN(t)$ indicates the monitoring process through which the outcome is observed. Covariates $Z(t)$ are only ``updated'' at times 0 and 2 and affect next outcomes and monitoring indicators.}\label{fig2}
 
\end{minipage}\vspace{0.4cm}

In the DGM in  Figures \ref{fig1} and \ref{fig2}, the visit rate depends on the covariate process but the fact of being monitored has no impact on the outcome or covariate processes. However, we postulate that our analysis of the marginal effect of antidepressants on BMI is such that previous monitoring affect the future monitoring path. For instance, the gap time (i.e.~the time since a last visit) is an endogenous covariate process which is modified by each subsequent monitoring indicator. The smoking status of each patient, or other health indicators, as well as drug prescriptions, could also be modified by a visit having taken place. When it is realistic to assume such an endogenous covariate process, and thus a ``joint'' process for the covariate, the monitoring and the outcome, dependencies between these processes may arise throughout the follow-up of a patient. This may mean that conditional on only the covariates measured most recently, the monitoring and the outcome processes are not independent. In particular, this may include situations where the monitoring path depends not only on the current covariates, but where it is also modified by what happened in the past (e.g., having had a physician visit yesterday makes my probability of visit today much lower). Effectively, previous monitoring indicators have interacted with covariates, such as the gap time, and the probability of visit on a given day may depend on the whole path the monitoring process went on (and all previous transitions). It may then be necessary to look at the monitoring path as a whole, and in particular to weight for the entire monitoring process -- and not simply the most recent monitoring event -- to ensure no unblocked paths between the intervention and outcome. 
 
 \begin{minipage}[t]{0.42\textwidth}
\centering
\begin{tikzpicture}[scale=0.6][%
-->,
shorten >=2pt,
>=stealth,
node distance=1cm,
pil/.style={
->,
thick,
shorten =2pt,}
]

\node (1) at (-0.6,0.4) {$ I(0) $};

 \node(3b) at (4,-2) {$Y(1)$};
\node(3c) at (8.5,-2) {$Y(2)$};
 
\node(5) at (2,1.4){\textcolor{black}{$\mathbf{K(0)}$}};
 
\draw[-{Latex[length=3mm]}] (1) to  (3b);
\draw[-{Latex[length=3mm]}] (1) to  (3c);
\draw[-{Latex[length=3mm]},black](5) to  (1);
 
 \draw[-{Latex[length=3mm]},black](5) to (3b);
\draw[-{Latex[length=3mm]},black](5) to (3c);
 
\node(30) at (-0.9, -3.8) {$Z(0)$};
\node(31) at (4, -4.1) {$Z(1)$};
 \node(32) at (9.3, -4.7) {$Z(2)$};

\draw[-{Latex[length=3mm]}](1) to (30);
\draw[-{Latex[length=3mm]}](1) to (31);
 \draw[-{Latex[length=3mm]}](1) to (32);

\node(11) at (4,-6) {\textcolor{black}{$dN(1)$}};
 \node(12) at (8.5,-6) {\textcolor{black}{$dN(2)$}};
 
 \node(s1) at (4.2,-8) {$*_1$};
\draw[-{Latex[length=3mm]}](30)  to (s1);
 \draw[-{Latex[length=3mm]}](11)  to (s1);
 \node(s2) at (7,-8) {$*_2$};
\draw[-{Latex[length=3mm]}](31)  to (s2);
 \draw[-{Latex[length=3mm]}](11)  to (s2);
 
\draw[-{Latex[length=3mm]}](1) to (11);
\draw[-{Latex[length=3mm]}](1) to (12);
\draw[-{Latex[length=3mm]}](30) to (11);

\draw[-{Latex[length=3mm]}](30) to (3b);

\draw[-{Latex[length=3mm]}](s1) to (12);
 \draw[-{Latex[length=3mm]}](s2) to (12);

\draw[-{Latex[length=3mm]}](s1) to (3c);
 \draw[-{Latex[length=3mm]}](s2) to (3c);

\end{tikzpicture} 
\captionsetup{font=footnotesize}

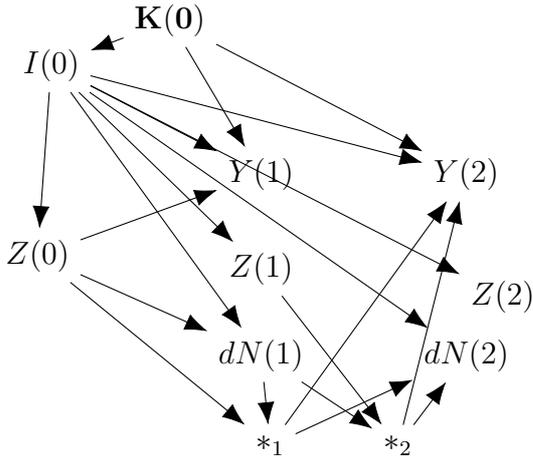
\captionof{figure}{Causal diagram for the third DGM (patient index $i$ removed) $I(0)$ is an intervention of interest whose marginal effect on a longitudinal outcome, $Y(t)$ -- assumed to be time-invariant -- is of interest. $\mathbf{K(0)}$ represent confounding variables, $Z(t)$ are mediators, and $dN(t)$ indicates the monitoring process through which the outcome is observed. Asterisks represent interactions between the covariates whose arrows point into it.}\label{fig3}
\end{minipage}\hspace{.8cm}
\begin{minipage}[t]{0.45\textwidth}
 \centering
\begin{tikzpicture}[scale=0.62][%
-->,
shorten >=2pt,
>=stealth,
node distance=1cm,
pil/.style={
->,
thick,
shorten =2pt,}
]

\node (1) at (-0.6,0.4) {$ I(0) $};

 \node(3b) at (4,-2) {$Y(1)$};
\node(3c) at (8.5,-2) {$Y(2)$};
 
\node(5) at (2,1.4){\textcolor{black}{$\mathbf{K(0)}$}};
 
\draw[-{Latex[length=3mm]}] (1) to  (3b);
\draw[-{Latex[length=3mm]}] (1) to  (3c);
\draw[-{Latex[length=3mm]},black](5) to  (1);
 
 \draw[-{Latex[length=3mm]},black](5) to (3b);
\draw[-{Latex[length=3mm]},black](5) to (3c);
 
\node(30) at (-0.9, -3.8) {$Z(0)$};
\node(31) at (4, -4.1) {$Z(1)$};
 \node(32) at (9.3, -4.7) {$Z(2)$};

\draw[-{Latex[length=3mm]}](1) to (30);
\draw[-{Latex[length=3mm]}](1) to (31);
 \draw[-{Latex[length=3mm]}](1) to (32);

\node(11) at (4,-6) {\textcolor{black}{$dN(1)$}};
 \node(12) at (8.5,-6) {\textcolor{black}{$dN(2)$}};
  
 \node(s1) at (4.2,-8) {$*_1$};
\draw[-{Latex[length=3mm]}](30)  to (s1);
 \draw[-{Latex[length=3mm]}](11)  to (s1);
 \node(s2) at (7,-8) {$*_2$};
\draw[-{Latex[length=3mm]}](31)  to (s2);
 \draw[-{Latex[length=3mm]}](11)  to (s2);
 
\draw[-{Latex[length=3mm]}](1) to (11);
\draw[-{Latex[length=3mm]}](1) to (12);
\draw[-{Latex[length=3mm]}](30) to (11);

\draw[-{Latex[length=3mm]}](30) to (3b);

\draw[-{Latex[length=3mm]}](s1) to (12);
 \draw[-{Latex[length=3mm]}](s2) to (12);

\draw[-{Latex[length=3mm]}](s1) to (3c);
 \draw[-{Latex[length=3mm]}](s2) to (3c);
 
 \draw[dashed, -{Latex[length=3mm]}](3b) to (3c);
 \draw[dashed, -{Latex[length=3mm]}](3b) to (12);
 
\end{tikzpicture} 
\captionsetup{font=footnotesize}

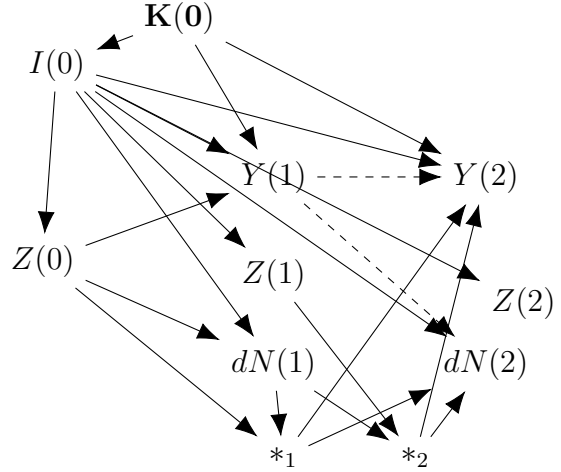
\captionof{figure}{Causal diagram for the fourth DGM (patient index $i$ removed) $I(0)$ is an intervention of interest whose marginal effect on a longitudinal outcome, $Y(t)$ -- assumed to be time-invariant -- is of interest. $\mathbf{K(0)}$ represent confounding variables, $Z(t)$ are mediators, and $dN(t)$ indicates the monitoring process through which the outcome is observed. Asterisks represent interactions between the covariates whose arrows point into it.  Dashed lines represent the new links added from the causal diagram presented in Figure \ref{fig3}.}\label{fig4}
 
\end{minipage}\vspace{0.4cm}

As an example of a scenario where these long-term dependencies may arise, we present a third DGM where the covariates are affected by the monitoring indicator through follow-up; practically, we implement this by including an interaction between the covariate and the visit indicator (Figure \ref{fig3}).  
In Figures \ref{fig3} and \ref{fig4}, we use a notation as in Moodie and Stephens (\citeyear{moodie2020comment}) and use the symbol $^*$ to refer to an interaction between the covariates whose arrows point into the $^*$. The interaction terms are necessarily deterministic, as an interaction term is solely determined by the respective variables that interact together. However, the monitoring indicators themselves are random, and an individual can transition to a visit or not on any given day.

In Figure \ref{fig3}, the covariate $Z(0)$ mediates the effect of $I(0)$ on $Y(1)$.  The covariate process $Z(\cdot)$ interacts with monitoring and whenever there is a new monitoring time $s$ (where $dN(s)=1$ for some $s>0$), the covariate process $Z(\cdot)$ is updated while still depending on the intervention at baseline. In that case, the selection bias due to outcome-dependent follow-up times cannot be adjusted for by using only the standard IIVW put forward by previous authors, as presented in equations (\ref{vp1}) or (\ref{vp2}) because a biasing path remains via the interactions between covariates $Z(\cdot)$ and the monitoring indicators $dN(\cdot)$. Using only the observed data and therefore conditioning on $dN(\cdot)$ opens colliders at the $dN(t)$, $\forall t>0$. After conditioning, one example biasing path from the intervention to the outcome is given by $I(0) \rightarrow Z(0) \rightarrow *_1 - dN(2) - *_2 \rightarrow Y(2)$. This path remains open even if we adjust for the last covariates observed $\mathbf{Z_i(l_i(t))}$ and for the intervention. Further, adjusting for the previous interaction term (between the last monitoring indicator and the most recent covariates observed) will not suffice, as this last interaction depends on the whole monitoring path (for instance, the interaction term for the previous monitoring indicator will be 0 if there was no visit yesterday, providing no adjustment for the previous non-null interaction term that occurred before yesterday).  
We graphically demonstrate some examples of biasing paths that arise after conditioning on the collider $dN(t)$ in the causal diagrams shown in Figures 1 to 4 along with a heuristic demonstration of why simpler weights do not always account properly for outcome-dependent monitoring times in Supplementary Material A.

 In Figure \ref{fig4}, the outcome process $Y(\cdot)$ also affects the monitoring rate for any given time in between the current monitoring time and the next monitoring time, as well as affecting the next outcome value. One consequence of this is that conditioning on $dN(2)$ opens a path between the intervention and the outcome through e.g.~the path $I(0) \rightarrow Z(0) \rightarrow *_1 - dN(2) -  Y(1) $.  Other biasing paths due to colliders $dN(t), t>0$, similar to those discussed for the third DGM, can also be found.

In the following section, we present our proposed inverse weighting method that can -- unlike previous approaches -- appropriately account for endogenous covariate-dependent monitoring processes as in the third and fourth described DGM above.

\subsection{A new weighting approach: Extension using the joint monitoring path}\label{hh}

To ensure that we break the dependence between the outcome and the monitoring processes in our estimation of the marginal effect of antidepressants on weight (or, similarly, in settings such as in the third and fourth DGMs depicted in Figures \ref{fig3} and \ref{fig4}), we propose an approach inspired by transition intensities and occupation probabilities used in the multistate models literature (Cook and Lawless, \citeyear{cook2018multistate}). In this approach, we  
effectively account for the full (observed) covariates history and the joint monitoring process.

The first step in our proposed approach is to model what we term a \textit{partial} conditional visit intensity at each point in time, which will represent the ``transition intensity'' to the state of being monitored (or a \textit{visit}). In this context, partial refers to the fact that we only condition on a subgroup of covariates measured in the past, and thus only require for these covariates to make subsequent monitoring indicators independent. We make the following assumption on monitoring indicators:
\begin{align}
\tag{I4} dN_i(t) \perp Y_i(t) | \mathbf{\mathcal{H}^o_i(t-)} \label{as3}
\end{align}
and further assume that only the subset $\left\{ \mathbf{Z_i(l_i(t))}, I_i(0), B_i(t-) \right\}$ of $\mathbf{\mathcal{H}^o_i(t-)}$ affecting the partial intensity at time $t$ are sufficient to break the dependence between subsequent monitoring indicators at each time $t$:
\begin{align}
\tag{I5} dN_i(t) \perp dN_i(t-) |   \mathbf{Z_i(l_i(t))}, I_i(0), B_i(t-),\label{as4}
\end{align}
where $dN_i(t-)$ is the last visit indicator observed prior to time $t$, and $B_i(t-)$, the last gap time. (In discrete time these would correspond to e.g.~$dN_i(t-1)$ and $B_i(t-1)$; note that these do not encompass the history of gap times and visits, only the values attached to the previous time unit.) Assumption (\ref{as4}) implies that given the previous gap time and the covariates observed at the previous visit time, the two subsequent monitoring indicators are independent of one another. This is a conditional Markov assumption for the monitoring process that allows us to decompose the process into a series of monitoring indicators. Note that $\mathbf{Z_i(l_i(t))}$ could contain many different kinds of predictors of the visit intensity, including mediators of the relationship between $I_i(0)$ and $Y_i(t)$ and functions of gap times or of time since cohort entry.   

  We quote Theorem 1 in Pearl (\citeyear{pearl2009causal}, pp. 110), which we rely on to compute the joint intensity of a given monitoring path: \vspace{0.2cm}\\
\noindent \textbf{Theorem 1} (The Causal Markov Condition). \textit{Any distribution generated by a Markovian model M can be factorized as:}
$$P(v_1, v_2, ..., v_n) = \prod_i P(v_i | pa_i)$$
\textit{ where }$V_1, V_2, ... V_n$ \textit{are the endogenous variables in M, and} $pa_i$ \textit{are (values of) the endogenous ``parents'' of $V_i$ in the causal diagram associated with M.} \vspace{0.3cm}

To model the partial conditional (visit transition) intensity at time $t$, based on assumption (\ref{as4}), we propose:
\begin{align}
\lambda_i(t | \mathbf{Z_i(l(t_i))},  I_i(0), B_i(t-)) = &  \lambda_0(B_i(t)) \exp(\gamma_I I_i(0) + \bm{\gamma_Z }\mathbf{Z_i(l_i(t))}), \label{modkp}
\end{align}

\noindent where $B_i(t)$ is a function of $B_i(t-)$ (which justifies the condition on that covariate in \ref{as4}).

To model a \textit{personalized} baseline intensity $\lambda_0(B_i(t))$, inspired by Zhu et al. (\citeyear{zhu2017estimation}), we propose a Breslow-type estimator (Cox, \citeyear{cox1972regression}), modified to be a function of the gap time since a last visit. For a gap time $B(t)$, it is given by
\begin{align}
\tag{S1} \hat{\lambda}_{0,1}(B(t))= \frac{ \sum^n_{i=1} \int_{s=0}^\tau  \mathbb{I}(dN_i(s)=1\cap B_i(s)=B(t))   }{ \sum_{i=1}^n \int_{s=0}^\tau \exp \left( \hat{\gamma}_I I_i(0)+ \bm{\hat{\gamma}_Z} \mathbf{Z_i(l_i(s))}\right) \mathbb{I}(dN_i(s)=1 \cap B_i(s)=B(t)) },\label{s1}
\end{align}
for $\mathbb{I}(\cdot)$ an indicator function. The logical statement $dN_i(s)=1 \cap  B_i(s)=B(t)$ means that patient $i$ both has a visit at time $s$ and that their gap time is $B_i(s)=B(t)$.
 
The intensities in (\ref{modkp}) are fitted using the Andersen and Gill model's main effects. As time is continuous, we compute the product integral of the transition intensities in (\ref{modkp}) to compute the probability of having a given monitoring path up to time $t$.  The product integral consists in the extension of the sum integral to the product (Gill and Johansen, \citeyear{gill1990survey}). A well-known estimator utilizing the product integral is the Kaplan-Meier estimator (\citeyear{kaplan1958nonparametric}). Here, transitions refer to those from the \textit{non-visit} to the \textit{visit} state, and vice-versa. For each time $t$, we assume the following \textit{simplified} transition matrix for individual $i$:
\begin{align*}
\begin{bmatrix}
1-\xi_i(t) \exp \left( \gamma_I I_i(0) + \bm{\gamma_Z} \mathbf{Z_i(l_i(t))}\right) \lambda_0(B_i(t)) dt & \xi_i(t) \exp \left( \gamma_I I_i(0) + \bm{\gamma_Z} \mathbf{Z_i(l_i(t))}\right) \lambda_0(B_i(t)) dt \\
1-\xi_i(t) \exp \left( \gamma_I I_i(0) + \bm{\gamma_Z} \mathbf{Z_i(l_i(t))}\right) \lambda_0(B_i(t)) dt & \xi_i(t) \exp \left( \gamma_I I_i(0) + \bm{\gamma_Z} \mathbf{Z_i(l_i(t))}\right) \lambda_0(B_i(t)) dt \\
\end{bmatrix},
\end{align*}
and, depending on which state was occupied at the very previous time unit, only some of these transitions will be non-null for each individual, at each time (this is why we call it \textit{simplified}; in reality, each element from the matrix above should be augmented with an indicator for the previous monitoring indicator, $\mathbb{I}(dN_i(t-))$, and the gap time and the covariates $\left\{I_i(0), \mathbf{Z_i(l_i(t))} \right\}$ will potentially be different on each row of the matrix, depending on whether or not there was a visit at time $t-$).
We rely on assumptions (\ref{as3}), (\ref{as4}), and \textbf{Theorem 1}, and take the product integral of the intensities for a given patient $i$; this leads to the probability of a given monitoring path conditional on the observed history of covariates, which is shown in (\ref{sw1}). The product symbol in (\ref{sw1}) refers to the product integral (as opposed to the standard product term).
\begin{align}
usw_i(t| \mathbf{\mathcal{H}^o_i(t-)}) &=\PRODI_{s=0}^t {\left\{ \xi_i(s) \exp \left( \gamma_I I_i(0) + \bm{\gamma_Z} \mathbf{Z_i(l_i(s))}\right) \lambda_0(B_i(s)) ds \right\} }^{\mathbb{I}(dN_i(s)=1) } \times\nonumber \\
&\hspace{1.3cm} {\left\{1-  \xi_i(s) \exp \left( \gamma_I I_i(0) + \bm{\gamma_Z} \mathbf{Z_i(l_i(s))}\right) \lambda_0(B_i(s)) ds \right\} }^{\mathbb{I}(dN_i(s)=0)}. \label{sw1}
\end{align}
 
\indent The weight in (\ref{sw1}) risks being highly variable. Furthermore, the inverse weight risks leading to extreme values, so that any estimator relying on it would consequently also have high variability.  To address this, we propose two alternative stabilizers to incorporate in (\ref{sw1}) and to be cumulated over each \textit{dt} time units. The first stabilizer is given by $\hat{\lambda}_{0,1}(B(t))$ as shown in equation (\ref{s1}), such that the baseline rate over each \textit{dt} time unit in (\ref{sw1}) cancels out. We compare this to a second proposed stabilizer (S2) that conserves the true effect of gap time from the fitted baseline rate. The second stabilizer uses a different estimator for the baseline rate that only depends on baseline covariates ($\mathbf{I(0)}$) and is given by
\begin{align}
\tag{S2} \hat{\lambda}_{0,2}(B(t))= \frac{ \sum^n_{i=1} \int_{s=0}^\tau  \mathbb{I}(dN_i(s)=1\cap B_i(s)=B(t))   }{ \sum_{i=1}^n \int_{s=0}^\tau \exp \left( \hat{\delta}_I I_i(0) \right) \mathbb{I}(dN_i(s)=1 \cap B_i(s)=B(t)) },
\end{align}
for $\delta$ the parameter in a proportional intensity model for monitoring times with the covariate $\mathbf{I(0)}$ as the only predictor. The second stabilizer (S2) does not depend on the endogenous covariate process $\mathbf{Z_i(l_i(\cdot))}$, but rather depends only on covariates measured at cohort entry. Unlike the stabilizer (S1) which results in a weight that does not adjust for the impact of the gap time on monitoring, the stabilizer (S2) may more completely account for structures where gap time itself affects the monitoring process.

Including the stabilizers $\lambda_{0,1}(B(t))$ or $\lambda_{0,2}(B(t))$ in the denominator in equation (\ref{sw1}) leads to the second proposed weight ($j \in  1, 2  $):
\begin{align}
sw_{i,j}(t| \mathbf{\mathcal{H}^o(t-)}) &=\PRODI_{s=0}^t \left( \frac{ \xi_i(s) \exp \left( \gamma_I I_i(0) +\bm{\gamma_Z} \mathbf{Z_i(l_i(s))}\right) \lambda_0(B_i(s)) ds  }{\lambda_{0,j}(B_i(s)) ds  }\right)^{\mathbb{I}(dN_i(s)=1) }\times \nonumber \\
&\hspace{1.4cm} \left(  \frac{ 1-  \xi_i(s) \exp \left( \gamma_I I_i(0) +\bm{\gamma_Z} \mathbf{Z_i(l_i(s))}\right) \lambda_0(B_i(s)) ds   }{ 1 - \lambda_{0,j}(B_i(s)) ds} \right) ^{\mathbb{I}(dN_i(s)=0)}. \label{sw2} 
\end{align}

By cumulating the intensity through time and by using its product as an inverse weight, we control (under stated assumptions) for the entire monitoring process conditional on the covariates' history. This weighting results in independence between the covariates and the monitoring process, so that their effect on the longitudinal outcome process can be estimated without bias. 
 
Once the weights are defined, similarly to Coulombe et al. (\citeyear{coulombe}), we use the following estimating equation for the marginal effect of intervention 
\begin{align}
\mathbb{E} \left( \int_0^{\tau}\frac{\mathbf{ Y(t)} -  [\bm{\beta_s}' \mathbf{S(t)}]}{\bm{w}\mathbf{(t|K(0))} \bm{sw_j}\mathbf{(t|  \mathcal{H}^o(t-))}  } \mathbf{dN(t)} \right) =\mathbf{0}, \label{eq}
\end{align}
where $\mathbf{S(t)}$ is a matrix containing a column of ones and a cubic spline basis to flexibly model time $t$ (together representing $\bm{\alpha}\mathbf{(t)}$), and the intervention variable $\mathbf{I(0)}$, and where $\beta \subset \bm{\beta_s}$ is the parameter of interest. That estimating equation is unbiased for the parameter of interest (proof given in Supplementary Material B).  
The weight $\mathbf{sw_j(t| \mathcal{H}^o(t-))}$ in (\ref{eq}) can be replaced by another with equivalent properties (balancing properties, positivity) such as $\mathbf{usw}\mathbf{(t|  \mathcal{H}^o(t-))}$, without modifying the rest of the estimating equation. The function $w_i(t|\mathbf{K(0)})$ can be estimated by using a correctly specified function of the covariates $\mathbf{K(0)}$ that breaks the dependence between the intervention and the confounders $\mathbf{K(0)}$. For instance, if the intervention and the confounders are time-fixed and assessed at time 0, an inverse probability of treatment weight can be used, and defined as the inverse of $w(t|\mathbf{K_i(0)}; \bm{\omega})= P( I_i(0)=1 | \mathbf{K_i(0)} ; \bm{\omega}) \mathbb{I}(I_i(0)=1) + (1- P( I_i(0)=1 | \mathbf{K_i(0)} ; \bm{\omega}) )\mathbb{I}(I_i(0)=0).$ That weight will account for imbalances due to confounders under the assumptions (\ref{v0}) and (\ref{p3}) if there is no conditioning on monitoring indicators, or under assumptions (\ref{v2}) and (\ref{p3}) if there is, and that the monitoring intensity is also taken into account.  
If the intervention is not time-fixed and if time-varying confounding exists that simultaneously acts as intermediate variables, methods such as presented in the seminal paper by Robins et al. (\citeyear{robins2000marginal}) can be used to recover balance.

For details on asymptoptic properties and how to compute a conservative asymptotic variance of the estimator for the marginal effect of intervention resulting from equation (\ref{eq}), we refer the reader to Supplementary Material C. In simulation studies, we use a nonparametric bootstrap to estimate the variance,  
re-sampling \textit{individuals} rather than observations to ensure within-person correlation of measures is preserved.

\section{Simulation studies}\label{se3}

We performed several simulation studies to assess the performance of the proposed weights to adjust for imbalances due to the informative monitoring process in settings similar to that of our research question which we answer using data from the CPRD. Our aim was to estimate the marginal effect of a binary (point) intervention on a continuous longitudinal outcome in contexts with confounding and informative monitoring times, where monitoring times were simulated in a sequential manner, dependent on previous information. 
 
 In the main study, we simulated for each patient $i$ three baseline confounders as $K_{1i}\sim \text{N}(1,1), K_{2i}\sim \text{Bernoulli}(0.55)$, and $K_{3i}\sim \text{N}(0,1)$. The intervention $I_i(t)$ was binary and time-fixed: $I_i\sim \text{Bernoulli}(p_{Ii})$ with $p_{Ii}=$ $\text{expit} \left( 0.5 + 0.8\hspace{0.02cm} K_{1i}+ 0.05\hspace{0.02cm} K_{2i} -1 \hspace{0.02cm}K_{3i}\right)$. One time-varying mediator $Z_i(\cdot)$ was generated, conditional on $I_i$. It was only updated whenever there was a new visit ($dN_i(\cdot)=1$), and was simulated as $Z_i(t)|I_i=1 \sim \text{N}(2,1)$ and $ Z_i(t)|I_i=0 \sim \text{N}(4,2^2)$
on those visit days. On other (non-visit) days, we denote the process by $Z_i(l_i(t))$, simply carrying forward the last observed value. Time was discretized over a grid of 0.01 units, from 0 to $\tau$. The intensity of monitoring at each time point over that grid was simulated as $\lambda_i(t| I_i,Z_i(l_i(t)))= 0.02 B_i(t) \exp \left( \gamma_1 I_i  + \gamma_2 Z_i(l_i(t)) \right)$. The outcome $Y_i(t)$ was generated according to $Y_i(t)= 0.2 B_i(t) + 1\hspace{0.05cm} I_i - 0.8\hspace{0.05cm} \left(Z_i(l_i(t)) - E\left[Z_i(l_i(t)) |I_i\right]\right) +  0.4\hspace{0.05cm} K_{1i}+ 0.05\hspace{0.05cm}K_{2i} -0.6 \hspace{0.05cm} K_{3i} +  \epsilon_i(t)$ with $ \epsilon_i(t) \sim \text{N}(0, 0.5^2)$. Monitoring times were drawn up until the maximum follow-up time $\tau$, which we fixed to $\tau=5$. Data were simulated to correspond to a study cohort of 500 patients. For each patient, the follow-up time was ``censored" (stopped) at time $C_i$, with $C_i\sim \text{Uniform}(\tau/2, \tau)$; the censoring was non-informative. A total of $1000$ replicate datasets were simulated for each simulation study scenario. More details on the simulation study can be found in Supplementary Material D.
 
We compared a naive ordinary least squares estimator that did not account for the confounding or the informative monitoring process ($\hat{\beta}_{LS}$) and an inverse probability of treatment weighted least squares estimator in which the treatment model was correctly specified but that did not account for the monitoring process  
($\hat{\beta}_{IPT}$) with four ``doubly weighted'' approaches that incorporated a correctly specified IPT weight and alternate versions of IIVW to account for the monitoring process. Specifically, these four estimators relied on intensity of visit weights 
that:
\begin{itemize} \setlength{\itemsep}{0pt}
\item  did not account for the full history of covariates but only for the last covariates observed, $I_i(0)$ and $Z_i(l_i(t))$ and for the correct baseline intensity through the following inverse weight
$ \hat{\lambda}_0(B_i(t)) \exp \left( \hat{\gamma_1} I_i  + \hat{\gamma_2} Z_i(l_i(t)) \right)$ (namely $\hat{\beta}_{IH}$),
\item used the novel inverse cumulated unstabilized weight ($\hat{\beta}_{USW}$),
\item used the cumulated stabilized weights $SW1$ ($\hat{\beta}_{SW1}$), or 
\item used the cumulated stabilized weights $SW2$ ($\hat{\beta}_{SW2}$)
\end{itemize}
where the latter two fully account for the full covariate process and its interaction with visit times but used different stabilizing strategies. 
 
The cumulated weights were censored at the respective 2.5 and 97.5th percentiles of their distribution for all three corresponding estimators.  

In four sensitivity analyses, we 1) fitted a constant intercept in the outcome model rather than a cubic spline basis for the effect of time; 2) varied the maximum follow-up time ($\tau=10$ rather than $\tau=5$); 3) changed the intercept function in the outcome from $0.2 B_i(t) $ to simply $0.02$; and 4) varied the endogenous process $Z(\cdot)$ such that its mean depended on the cumulative number of previous visits, as: $Z_i(t)|I_i=1 \sim \text{N}(2 + 0.2 \hspace{0.1cm} \int_0^{t-} dN_i(s),1)$ and $ Z_i(t)|I_i=0 \sim \text{N}(4+ 0.2 \hspace{0.1cm}\int_0^{t-} dN_i(s),2^2)$ updated on visit days, with the rest unchanged.
The results for the sensitivity analyses are presented in Supplementary Material F. 

Table \ref{tab:res1} shows the results of the main simulation study in terms of empirical absolute bias and bootstrap variance. The Andersen and Gill model consistently provided unbiased estimates of the parameters $\bm{\gamma}$ in the monitoring model in all these scenarios, and average numbers of visits varied between 1.9 and 7.1 (Supplementary Material E). The Breslow-type estimator used to estimate $\lambda_0(B_i(t))$ provided consistent estimates (results not shown).  

In the main analysis, the least squares estimator was biased (bias ranging between 0.35-0.73). Accounting for the confounding improved the performance of the estimator, as seen with the absolute bias of $\hat{\beta}_{IPT}$ (range between 0.01 and 0.37). However, both the IPW estimator and the doubly weighted estimator $\hat{\beta}_{IH}$ remained biased, with the range of bias depending on the strength of the effect of covariates in the visit process (as controlled by the $\bm{\gamma}$s). This supports our claim that existing approaches do not fully adjust for a visit process in settings with endogeneity and long-term dependencies.

With respect to the cumulated weights, only  
$\hat{\beta}_{SW2}$ consistently provided unbiased estimates. Its variance was also relatively small, compared with the other simpler estimators. The unstabilized weighted estimator $\hat{\beta}_{USW}$ was highly variable, and censoring its cumulated intensity weights at the 2.5th and 97.5 percentiles did not reduce its variance to a satisfactory degree (Table \ref{tab:res1}). 
This could be due to the sample size, or the number of simulations conducted. We present the absolute bias; the empirical bias varied between negative and positive, so the direction of the bias was not systematic. For the first of the stabilized weighted estimators, $\hat{\beta}_{SW1}$, we hypothesize that the weights might not adequately account for the effect of time since the baseline rate canceled out after stabilization. In fact, we observed that the estimator $\hat{\beta}_{SW1}$ performed slightly better than $\hat{\beta}_{IH}$ in general. This may be due to its adjustment for part of the endogeneity/dependence due to covariates since the data were generated such that  
$Z(\cdot)$ was  
simulated according to a random Normal variable with a mean that did not vary across follow-up. Although on average, the process remained centered around the same value, adjusting for the process $Z(\cdot)$ in $\hat{\beta}_{SW1}$ nonetheless accounted for the variation of $Z(\cdot)$ around its mean.  

In sensitivity analyses (Supplementary Material F), we found very similar results, with $\hat{\beta}_{SW2}$ performing better across the board. All results remained similar when using a constant intercept as compared to a more flexible intercept with a cubic spline basis, as well as when increasing the maximum follow-up time, $\tau$, to 10. When changing the endogenous covariate distribution that affected the monitoring and the outcome processes to make it dependent on the cumulative number of previous visit times, bias was not,  in general, much greater. In the fourth sensitivity analysis where we changed the intercept function to a constant in the definition of the outcome, similar results were observed.
\begin{table}[H]
 \begin{center}
\caption{Main analysis: Mean absolute bias and bootstrap variance for the estimators compared (1000 simulations, $n=500$ patients).}
\begin{tabular}{ ccccccccc c cc c }
 \hline \hline
$\bm{\gamma}$ &   \multicolumn{6}{c} {Mean absolute bias of the estimator}  & \multicolumn{6}{c} {Bootstrap variance of the estimator}   \\
  &     $\hat{\beta}_{LS}$&$\hat{\beta}_{IPT}$&$\hat{\beta}_{IH}$&$\hat{\beta}_{USW}$  & $\hat{\beta}_{SW1}$   & $\hat{\beta}_{SW2}$ &  $\hat{\beta}_{LS}$&$\hat{\beta}_{IPT}$&$\hat{\beta}_{IH}$&$\hat{\beta}_{USW}$  & $\hat{\beta}_{SW1}$   & $\hat{\beta}_{SW2}$\\  
\hline
 -0.3; 0.1& 0.35&0.37&0.12&0.04&0.14&0.03  &0.03 &0.07 &0.16 &1.56 &0.13 &0.08\\
 -0.2; 0.2& 0.49 &0.24&0.11&0.00&0.09&0.01 & 0.03&0.06 &0.11 &1.23 &0.08 &0.06\\
 -0.1; 0.2& 0.64 &0.08&0.09&0.14&0.00&0.02& 0.02 &0.05 &0.09 &1.11 &0.06 &0.05 \\
-0.1; -0.3& 0.69&0.01&0.11&0.07&0.01&0.03 &0.02& 0.05& 0.08&0.59 &0.06 &0.05\\
0; 0& 0.73 &0.01&0.03&0.16&0.01&0.01 & 0.02 &0.05 &0.09 &0.56 &0.05 & 0.05\\
0.1; -0.3& 0.69 &0.03&0.11&0.19&0.04&0.02 & 0.02 &0.04 & 0.06&0.40 &0.05 &0.05\\
0.2; -0.2& 0.64 &0.12&0.26&0.18&0.19&0.02 &0.02 &0.03 &0.06 &0.36 &0.05 &0.05 \\
0.3; 0.2&  0.67&0.08 &0.34 &0.24&0.30&0.05&  0.02& 0.03&0.06 &0.26 &0.07 &0.05 \\   
\hline
 \end{tabular}
 \label{tab:res1}
\end{center}
\end{table}
The results for the variance were as expected: the use of unstabilized cumulated weight in $\hat{\beta}_{USW}$ led to a large variance, while the variance of other estimators remained low. Stabilizing the cumulated weights in $\hat{\beta}_{SW1}$ and $\hat{\beta}_{SW2}$ led to smaller variance, in general, than the more classical (and, in these settings, biased) estimator that accounted for the visit process, $\hat{\beta}_{IH}$. A comparison of the empirical and the bootstrap variances of all estimators compared in the main analysis can be found in Supplementary Material G; the bootstrap variance was slightly greater than the empirical variance, in general.

\section{Comparison of the effect of citalopram and fluoxetine on BMI using the CPRD data}\label{analysis}

We use the proposed methodology to estimate the marginal effect of citalopram and fluoxetine on BMI, relying on data from the Clinical Practice Research Datalink in the United Kingdom (UK). The CPRD is one of the largest primary care databases of de-identified data. It contains data from more than 13 million patients treated in general practices from across the UK, including demographics, anthropometric measurements such as BMI, lifestyle factors, all prescriptions issued by general practitioners (recorded according to the British National Formulary), and medical diagnoses (coded using the Read Classification System). The CPRD data were linked with the Hospital Episode Statistics (HES) repository and the Office for National Statistics (ONS) mortality database. These provided access to further patients information on related diagnoses for each hospital stay (coded using the International Classification of Diseases version 10), and dates of death.

Our study protocol was approved by the Independent Scientific Advisory Committee of the UK Clinical Practice Research Datalink (protocol number $19\_017R$) and the Research Ethics Committee of the Jewish General Hospital (Montreal, Quebec, Canada).  
 
We defined a cohort of adult new users of citalopram or fluoxetine who had a confirmed diagnosis of depression in the year prior to initiation. To be included, patients had to initiate their treatment for one of the two study drugs between April 1st, 1998, and  December 31st, 2017. The final cohort comprised 246,503 patients (56\% citalopram new users); see Coulombe et al. (\citeyear{coulombe2}) for details of the cohort construction. Patients were followed until a first code for pregnancy, treatment discontinuation for citalopram or fluoxetine, switch to any other antidepressant drug, end of CPRD coverage, administrative end of study (December 31st, 2017), death, or when reaching a maximum follow-up time set to 18 months, whichever happened first. Patients were considered continuously exposed if a subsequent prescription for the initiating drug was issued less than 30 days after the end of the duration of the last prescription for the corresponding drug. 
  
Over the follow-up period, we collected any data on BMI and considered as a monitoring time any day when BMI was recorded by the general practitioner. Other days when either 1) no physician or hospital visit occurred; or 2) such visit occurred but no BMI was measured and recorded, were considered to be the same (i.e. \textit{no monitoring}); we did not model them any differently.  

The intervention (citalopram or fluoxetine), age, sex, and Index of Multiple Deprivation (IMD) were defined at baseline and were included as predictors in the BMI monitoring model. The IMD is a measure of relative deprivation for small areas in England  (Deas et al., \citeyear{deas2003measuring}); it was used as a proxy for socioeconomic status. We assumed that certain time-varying covariates were potentially modified by a visit and could affect the next BMI outcome and the timing of the next visit. These covariates were defined for each day during follow-up, and included the smoking status, diabetes or antidiabetic drug use, alcohol abuse, a diagnosis for anxiety or generalized anxiety disorder (GAD), other psychiatric diseases (including autism spectrum disorder, obsessive compulsive disorder, and schizophrenia), the number of hospitalisations in the previous month, and benzodiazepine drug, lipid lowering drug, or antipsychotic drug use (all three considered separately). We fitted a proportional intensity model as a function of all covariates mentioned above for the visit model. The visit intensity was modelled as a function of the gap time $B_i(t)$, accounted for via the baseline intensity $\lambda_0(B_i(t))$, and all predictors in the model. It was used to compute the different weights and corresponding estimators described in Section 2 and compared in Section 3. For every time unit (one day), we obtained an estimate of the visit intensity.

We assumed that the relationship between the intervention (citalopram or fluoxetine) and BMI was potentially confounded by a set of covariates measured at baseline. These included age, sex, the IMD, smoking status, diabetes, alcohol abuse, anxiety or GAD, other psychiatric diseases, the number of previous hospitalisations, as well as the use of lipid lowering therapy, benzodiazepine drugs, or antipsychotic drugs. We fitted a logistic regression model to predict the intervention at baseline (citalopram or fluoxetine) and included the confounder variables mentioned above as predictors in the model. The fit led to a propensity score that was used in an inverse probability of treatment weight, to address confounding.  

For the monitoring and the exposure models mentioned above, covariates defined once at baseline were defined in the same way as in Coulombe et al. (\citeyear{coulombe2}) (any medication use was measured in the year prior to cohort entry, and comorbidities using any data recorded prior to cohort entry). For time-varying covariates, we used different definitions; we considered a patient exposed to a given medication for the duration of a prescription and unexposed otherwise. After any first diagnosis for a chronic disease during follow-up (diabetes, alcohol abuse, anxiety or GAD and other psychiatric diseases), a patient was considered as having the disease for the remainder of the follow-up. At any given time during follow-up, the time-varying smoking status was defined using the very last code recorded for smoking. Contrary to the simulation studies, visits to general practices could occur even in between times when BMI was monitored, such that patients' characteristics other than BMI were being updated in between BMI monitoring times, regardless of whether BMI was monitored or not on those days.

The (different versions of the) inverse weight for the monitoring, along with the inverse probability of treatment weight were used to perform a weighted regression for the outcome, which included a cubic spline basis to model the effect of time since a last visit (``gap time'') on the mean outcome, and the intervention of interest.  
 For the proposed estimators ($\hat{\beta}_{USW}$, $\hat{\beta}_{SW1}$, and $\hat{\beta}_{SW2}$), the cumulated weights were truncated using the 2.5th and 97.5th percentiles as in simulation studies.  
 There were a few differences across the two treatment groups at baseline (Supplementary Material H).  
 The most important difference was in the prevalence of anxiety or GAD (30.5\% in citalopram users vs 22.1\%). The average BMI at cohort entry was 26.8 in both treatment groups. During follow-up, the average BMI in citalopram users shifted to 28.3 (median 27.2), as compared to 28.9 (median 27.6) in fluoxetine users (results not shown), indicating a slightly different shift between the two groups as compared to baseline.
  
  In the citalopram group, we found an average of 0.28 visit per patient over the entire follow-up, with an average follow-up time of 0.45 years, yielding a crude visit (or BMI monitoring) rate of 0.62 visit per year.  
 In the fluoxetine users, the average number of visits per year was 0.25, with an average follow-up time of 0.40 years, yielding the same crude monitoring rate of 0.62 visit/year.   
  \begin{table}[H]
 \begin{center}
\caption{Rate ratios from the visit intensity model, Clinical Practice Research Datalink, UK, 1998-2017}
\begin{tabular}{ lcccc }
 \hline \hline
Variable &Rate Ratio& 95\% CI\\ \hline
 
  Citalopram (Ref.: Fluoxetine) &0.95& 0.93, 0.96* \\
    Age at baseline       &1.00& 0.99, 1.00\textcolor{white}{*} \\
  Sex (Ref.: Female)     &0.77& 0.76, 0.78* \\
  IMD at baseline       &1.06& 1.05, 1.06* \\
Smoking (Ref.: Never)  & & \\  
$\hspace{0.5cm}$Ever &0.92& 0.91, 0.94* \\
$\hspace{0.5cm}$Missing     &0.23& 0.23, 0.23* \\
 Diabetes     &2.07& 2.02, 2.12* \\
 Alcohol abuse  &1.14& 1.07, 1.22* \\
      Anxiety or GAD &1.00& 0.98, 1.03\textcolor{white}{*} \ \\
      Psychiatric diagnosis    &1.03& 0.93, 1.13\textcolor{white}{*} \\\
  Number of hospitalisations in prior month         &0.98& 0.96, 1.01\textcolor{white}{*} \ \\
  Antipsychotic drugs         &1.12& 1.07, 1.18* \\
   Benzodiazepine drugs         &1.20& 1.16, 1.23* \\
  Lipid lowering drugs         &1.21& 1.18, 1.25* \\
 \hline
\end{tabular}
 \label{tab:res3}
 
 \scriptsize{Abbreviations: IMD, Index of multiple deprivation; GAD, Generalized anxiety disorder.}\\
 \scriptsize{* Confidence interval does not contain 1.}
  \end{center}
 \end{table}
 Table \ref{tab:res3} shows the adjusted rate ratios for monitoring estimated from the Andersen and Gill model, for all covariates in the multivariate monitoring intensity model, along with corresponding 95\% confidence intervals (CIs) obtained from the model. 
 
We found that males, citalopram users (as opposed to fluoxetine), and those with a previous record of smoking or no smoking information at all (as opposed to non-smokers) were less likely to have their weight recorded. A greater IMD, alcohol abuse, diabetes as well as the use of antipsychotic, benzodiazepine, or lipid lowering drugs were all associated with an increase in the rate of BMI monitoring. It is unclear whether these time-dependent covariates could lead to long-term biasing dependencies between the monitoring and the BMI processes, as both of these processes vary in time as a function of these covariates. Estimating the marginal effect of antidepressants on BMI after re-weighting only for a point inverse intensity of visit weight, as opposed to a cumulated weight, could provide different estimates if the cumulated weight indeed provides further adjustment (e.g.,~if the probability of visit at time $t$ in individual $i$ is only proportional to that of another individual $j$ on their set of covariates, when accounting for the full history of intensities).

When we incorporated the same set of covariates in an outcome model for the continuous outcome BMI, we found that being older, male, using citalopram (as opposed to fluoxetine), alcohol abuse, smoking, a greater number of hospitalisations in the previous month, and the use of benzodiazepine drugs were statistically significantly associated with a lower BMI (Supplementary Material I). On the other hand, a greater IMD, diabetes, and the use of lipid lowering therapy were significantly associated with a greater BMI. Of note, several covariates were associated with both monitoring and BMI value (Tables \ref{tab:res3} and Supplementary Material I). These covariates may have induced selection bias due to outcome-dependent monitoring times.  In particular, diabetes was strongly associated with both the monitoring rate, and the outcome value.  Previous literature suggested that diabetes is a mediator of antidepressant drugs' effect on weight, as antidepressant therapy is associated with poor glycemic control (see e.g. Gagnon et al., \citeyear{gagnon2018impact}).

In Table \ref{tab:res4} we present the estimates for the marginal effect of citalopram (as compared to fluoxetine) on BMI, for each of the six estimators we compare, along with the 95\% robust CIs. Using the two cumulated weights and accounting for the possibility for long-term dependencies brought the estimates further away from the null (coefficients around -0.61 to -0.73) as compared to the more standard inverse intensity of visit weighted estimator ($\hat{\beta}_{IH}$, coefficient  -0.40, 95\% CI: -0.58, -0.22).

 \begin{table}[H]
 \begin{center}
\caption{Comparison of six estimators for the marginal effect of citalopram (as opposed to fluoxetine) on BMI, Clinical Practice Research Datalink, UK, 1998-2017 }
\begin{tabular}{ l c}
 \hline \hline
Estimator & Estimate (Robust 95\% CI)\\ \hline
 
$\hat{\beta}_{LS}$ & -0.58 (-0.70, -0.46) \\
$\hat{\beta}_{IPT}$ & -0.65 (-0.78, -0.53) \\
$\hat{\beta}_{IH}$ & -0.40 (-0.58, -0.22)  \\
$\hat{\beta}_{USW}$ & -0.73 (-1.03, -0.44) \\
$\hat{\beta}_{SW1}$ & -0.62 (-0.78, -0.46) \\
 $\hat{\beta}_{SW2}$ & -0.61 (-0.76, -0.46) \\
   \hline
\end{tabular}
 \label{tab:res4}

 \end{center}
 \end{table}

Using our proposed cumulated intensity weight ($\hat{\beta}_{SW2}$), as opposed to a simpler weight ($\hat{\beta}_{IH}$), resulted in a change of approximately 50\% in the point estimate of the marginal effect of citalopram, in this study.  The difference could indicate that long-term dependencies between the covariate, the monitoring and the outcome processes indeed exist. However, all estimators and the associated CIs suggest the same conclusion: that citalopram leads to less weight gain than fluoxetine. The effect of the two study drugs on BMI, as well as the difference in effects, remains modest, although we remind the reader that the follow-up time was relatively short.

\section{Discussion}
 
In studies using electronic health records or administrative data, \textit{when} patients' information is recorded often depends on patients' characteristics. The informative nature of monitoring times may be associated with biasing paths between the intervention and the outcome under study, as monitoring can be a source of selection bias. In our analysis of the CPRD data, several patient characteristics were associated both with the monitoring rate and the BMI values, potentially inducing selection bias in the estimation. No previous studies have estimated the marginal effect of citalopram and fluoxetine on BMI while accounting for this type of bias (along with confounding).

It can be unclear whether the dependence between the monitoring process and the BMI process extends beyond the last covariates observed. This work proposes some first insights into this. We proposed and demonstrated  a new methodology to address that dependence by accounting for the potential for longitudinal collider-stratification bias due to an endogenous covariate process. In the CPRD data, we estimated the marginal effect of prescribing citalopram versus fluoxetine on BMI. The proposed weights did provide different estimates for that effect, as compared to the more standard IIVW that does not fully account for the covariate-dependent monitoring path. However, the differences were clinically modest. In general, a comparison of our proposed estimators and other simpler estimators that are not cumulated over time could provide indications of whether long-term dependencies are present if estimates differ substantially across approaches. In simulation studies, the proposed stabilized cumulated weighted estimator ($\hat{\beta}_{SW2}$) was the only estimator to be consistently unbiased for the marginal effect of intervention across all scenarios with endogeneity. The stabilization yielded more efficient estimation. 

The proposed weights that account for the covariate-driven monitoring times are similar to that used in marginal structural models to address confounding in longitudinal treatment sequences (Robins, Hernàn and Brumback, \citeyear{robins2000marginal}) and to the calculation of stage occupation probability in the multistate models literature for settings with continuous time (Cook and Lawless, \citeyear{cook2018multistate}) but they tackle imbalances due to the monitoring process rather than that due to confounding factors. We combined them with inverse probability of treatment weights to account simultaneously for confounding. Together, these weights create a pseudo-population in which the monitoring and the BMI processes are independent, and in which the two antidepressant groups are exchangeable, so as to permit inference about the marginal effect of both antidepressants on BMI. This study is the first to assess the marginal effect of citalopram and fluoxetine on BMI while considering that monitoring times are driven by an endogenous covariate process, and the first to propose a methodology for it. Another key strength of this work is our simplifying assumption on the covariate process; it allows for the covariates to be assessed occasionally (at monitoring times) for the proposed weights to break the dependence between the monitoring and BMI processes. Further, our weights allow for the analyst to account for the effect of mediators of the antidepressant-BMI relationship that affect monitoring times without blocking the total effect of interest.  

The proposed estimators rely on important assumptions. In particular, the model for the visit intensity function must be correctly specified. If covariates ``occurring'' or being updated in between monitoring times induce dependency between the monitoring and the BMI processes, and that they are not accounted for in the intensity model, then the proposed estimator could be biased. We also made the strong assumption of positivity of the monitoring process. When it is unrealistic, this assumption could be circumvented by smoothing the intensity function, e.g.~by coarsening the monitoring indicators. Moreover, the proposed estimators rely on the standard identifiability assumptions in causal inference. For instance, the presupposition of conditional exchangeability assumes that we measured all potential confounders for the relationship between the antidepressants and BMI, an assumption that cannot be verified in practice. Sensitivity analyses were proposed to assess the extent to which unmeasured confounding can affect the estimator for the marginal effect of exposure (see e.g. Streeter et al., \citeyear{streeter2017adjusting} for a review of methods, for longitudinal settings). The assumption about the positivity of treatment, on the other hand, could be implausible in other settings, but in this work, it is unlikely to be violated as citalopram and fluoxetine are often prescribed interchangeably in patients with depression. %
In other situations where this assumption is not plausible, patients who have no chance of receiving some treatment options could be removed (or part of their person-time in the study) at the cost of reduced generalizability.

Although citalopram and fluoxetine are front-line treatments for depression and hence very commonly prescribed, side effects remain a significant challenge for users. In particular, weight gain may be substantial and so it is of considerable interest to use data from a general population to determine the impact of these drugs and to see whether one might lead to a lower burden of this particular side effect.
In the first analysis of  electronic health records  data from a large, population-based sample, we have found that citalopram leads to (modestly) less weight gain than fluoxetine, after adjusting for biases due to confounding and the covariate-induced visit process. These findings must be interpreted with caution as, of course, clinical decisions must balance a number of additional factors. Nevertheless, this analysis serves as an important model for considerations that are required when working with EHR data.

\begin{center}
{\large\bf Supplementary material}
\end{center}
The following supplementary material is available at the very end of this document, after the references:
\begin{description}
\item[\textbf{Suppl. Material A:}] Causal diagrams and biasing paths due to the monitoring process
\item[\textbf{Suppl. Material B:}] Estimating equation for the marginal effect of treatment on a continuous longitudinal outcome 
\item[\textbf{Suppl. Material C:}] Asymptotic properties of the proposed estimator
\item[\textbf{Suppl. Material D:}] Details of the simulation studies
\item[\textbf{Suppl. Material E:}] Results of the main simulation study, including the average number of visits and estimated parameters in the visit model
\item[\textbf{Suppl. Material F:}] Results of all sensitivity analyses
\item[\textbf{Suppl. Material G:}] Comparison of the bootstrap and the empirical variance of the estimators
\item[\textbf{Suppl. Material H:}] Table of baseline characteristics stratified by intervention group in the CPRD  
\item[\textbf{Suppl. Material I:}] Multivariate outcome model in the analysis of the CPRD  
\end{description}

\section*{Acknowledgments}

This work is supported by a doctoral scholarship from the Natural Sciences and Engineering Research Council (NSERC) of Canada (Ref.~401223940) to author JC. EEMM acknowledges support from a Discovery Grant from NSERC and a chercheur-boursier career award from the Fonds de recherche du Québec--Santé. RWP acknowledges support from a Discovery Grant from NSERC and a Foundation Scheme Grant from CIHR.

\makeatother
\bibliographystyle{apalike}  
\bibliography{arxiv.bib}        

\begin{thebibliography}{}

\bibitem[Andersen and Gill, 1982]{andersen1982cox}
Andersen, P.~K. and Gill, R.~D. (1982).
\newblock Cox's regression model for counting processes: a large sample study.
\newblock {\em The Annals of Statistics}, 10(4):1100--1120.

\bibitem[Bůžková and Lumley, 2009]{buuvzkova2009semiparametric}
Bůžková, P. and Lumley, T. (2009).
\newblock Semiparametric modeling of repeated measurements under
  outcome-dependent follow-up.
\newblock {\em Statistics in Medicine}, 28(6):987--1003.

\bibitem[Cook and Lawless, 2007]{cook2007statistical}
Cook, R.~J. and Lawless, J. (2007).
\newblock {\em The statistical analysis of recurrent events}.
\newblock New York: Springer Science \& Business Media.

\bibitem[Cook and Lawless, 2018]{cook2018multistate}
Cook, R.~J. and Lawless, J.~F. (2018).
\newblock {\em Multistate models for the analysis of life history data}.
\newblock Boca Raton: CRC Press.

\bibitem[Coulombe et~al., 2020a]{coulombe}
Coulombe, J., Moodie, E. E.~M., and Platt, R.~W. (2020a).
\newblock Weighted regression analysis to correct for informative monitoring
  times and confounders in longitudinal studies.
\newblock {\em Biometrics}.
\newblock Forthcoming.

\bibitem[Coulombe et~al., 2020b]{coulombe2}
Coulombe, J., Moodie, E. E.~M., Shortreed, S., and Renoux, C. (2020b).
\newblock Can the risk of severe depression-related outcomes be reduced by
  tailoring the antidepressant therapy to patient characteristics?
\newblock {\em American Journal of Epidemiology}.
\newblock Forthcoming.

\bibitem[Cox, 1972]{cox1972regression}
Cox, D.~R. (1972).
\newblock Regression models and life-tables.
\newblock {\em Journal of the Royal Statistical Society: Series B
  (Methodological)}, 34(2):187--202.

\bibitem[De~las Cuevas et~al., 2014]{de2014risk}
De~las Cuevas, C., Pe{\~n}ate, W., and Sanz, E.~J. (2014).
\newblock Risk factors for non-adherence to antidepressant treatment in
  patients with mood disorders.
\newblock {\em European Journal of Clinical Pharmacology}, 70(1):89--98.

\bibitem[Deas et~al., 2003]{deas2003measuring}
Deas, I., Robson, B., Wong, C., and Bradford, M. (2003).
\newblock Measuring neighbourhood deprivation: a critique of the index of
  multiple deprivation.
\newblock {\em Environment and Planning C: Government and Policy},
  21(6):883--903.

\bibitem[Gagnon et~al., 2018]{gagnon2018impact}
Gagnon, J., Lussier, M.-T., MacGibbon, B., Daskalopoulou, S.~S., and Bartlett,
  G. (2018).
\newblock The impact of antidepressant therapy on glycemic control in canadian
  primary care patients with diabetes mellitus.
\newblock {\em Frontiers in Nutrition}, 5:47.

\bibitem[Gill and Johansen, 1990]{gill1990survey}
Gill, R.~D. and Johansen, S. (1990).
\newblock A survey of product-integration with a view toward application in
  survival analysis.
\newblock {\em The Annals of Statistics}, 18(4):1501--1555.

\bibitem[Greenland, 2003]{greenland2003quantifying}
Greenland, S. (2003).
\newblock Quantifying biases in causal models: classical confounding vs
  collider-stratification bias.
\newblock {\em Epidemiology}, 14(3):300--306.

\bibitem[Herrett et~al., 2015]{herrett2015data}
Herrett, E., Gallagher, A.~M., Bhaskaran, K., Forbes, H., Mathur, R., Van~Staa,
  T., and Smeeth, L. (2015).
\newblock Data resource profile: clinical practice research datalink {(CPRD)}.
\newblock {\em International Journal of Epidemiology}, 44(3):827--836.

\bibitem[Kaplan and Meier, 1958]{kaplan1958nonparametric}
Kaplan, E.~L. and Meier, P. (1958).
\newblock Nonparametric estimation from incomplete observations.
\newblock {\em Journal of the American statistical association},
  53(282):457--481.

\bibitem[Liang et~al., 2009]{liang2009joint}
Liang, Y., Lu, W., and Ying, Z. (2009).
\newblock Joint modeling and analysis of longitudinal data with informative
  observation times.
\newblock {\em Biometrics}, 65(2):377--384.

\bibitem[Lin et~al., 2000]{lin2000semiparametric}
Lin, D.~Y., Wei, L.-J., Yang, I., and Ying, Z. (2000).
\newblock Semiparametric regression for the mean and rate functions of
  recurrent events.
\newblock {\em Journal of the Royal Statistical Society: Series B (Statistical
  Methodology)}, 62(4):711--730.

\bibitem[Lin et~al., 2004]{lin2004analysis}
Lin, H., Scharfstein, D.~O., and Rosenheck, R.~A. (2004).
\newblock Analysis of longitudinal data with irregular, outcome-dependent
  follow-up.
\newblock {\em Journal of the Royal Statistical Society: Series B (Statistical
  Methodology)}, 66(3):791--813.

\bibitem[Lindsey, 2004]{lindsey2004statistical}
Lindsey, J.~K. (2004).
\newblock {\em Statistical Analysis of Stochastic Processes in Time},
  volume~14.
\newblock Cambridge: Cambridge University Press.

\bibitem[Lipsitz et~al., 2002]{lipsitz2002parameter}
Lipsitz, S.~R., Fitzmaurice, G.~M., Ibrahim, J.~G., Gelber, R., and Lipshultz,
  S. (2002).
\newblock Parameter estimation in longitudinal studies with outcome-dependent
  follow-up.
\newblock {\em Biometrics}, 58(3):621--630.

\bibitem[Moodie and Stephens, 2020]{moodie2020comment}
Moodie, E. E.~M. and Stephens, D.~A. (2020).
\newblock Comment: clarifying endogeneous data structures and consequent
  modelling choices using causal graphs.
\newblock {\em Statistical Science}, 35(3):391--393.

\bibitem[Neyman, 1923]{neyman1923application}
Neyman, J.~S. (1923).
\newblock On the application of probability theory to agricultural experiments.
  {E}ssay on principles. {S}ection 9 (translation published in 1990).
\newblock {\em Statistical Science}, 5(4):472--480.

\bibitem[Pearl, 2009]{pearl2009causal}
Pearl, J. (2009).
\newblock Causal inference in statistics: an overview.
\newblock {\em Statistics Surveys}, 3:96--146.

\bibitem[Pullenayegum, 2016]{pullenayegum2016multiple}
Pullenayegum, E.~M. (2016).
\newblock Multiple outputation for the analysis of longitudinal data subject to
  irregular observation.
\newblock {\em Statistics in Medicine}, 35(11):1800--1818.

\bibitem[Pullenayegum and Lim, 2016]{pullenayegum2016longitudinal}
Pullenayegum, E.~M. and Lim, L.~S. (2016).
\newblock Longitudinal data subject to irregular observation: a review of
  methods with a focus on visit processes, assumptions, and study design.
\newblock {\em Statistical Methods in Medical Research}, 25(6):2992--3014.

\bibitem[Robins et~al., 2000]{robins2000marginal}
Robins, J.~M., Hern\`an, M.~A., and Brumback, B. (2000).
\newblock Marginal structural models and causal inference in epidemiology.
\newblock {\em Epidemiology}, 11(5):550--560.

\bibitem[Rosenbaum and Rubin, 1983]{rosenbaum1983central}
Rosenbaum, P.~R. and Rubin, D.~B. (1983).
\newblock The central role of the propensity score in observational studies for
  causal effects.
\newblock {\em Biometrika}, 70(1):41--55.

\bibitem[Rubin, 1974]{rubin1974estimating}
Rubin, D.~B. (1974).
\newblock Estimating causal effects of treatments in randomized and
  nonrandomized studies.
\newblock {\em Journal of Educational Psychology}, 66(5):688.

\bibitem[Serretti et~al., 2010]{serretti2010antidepressants}
Serretti, A., Mandelli, L., Laura, M., et~al. (2010).
\newblock Antidepressants and body weight: a comprehensive review and
  meta-analysis.
\newblock {\em The Journal of Clinical Psychiatry}, 71(10):1259--1272.

\bibitem[Streeter et~al., 2017]{streeter2017adjusting}
Streeter, A.~J., Lin, N.~X., Crathorne, L., Haasova, M., Hyde, C., Melzer, D.,
  and Henley, W.~E. (2017).
\newblock Adjusting for unmeasured confounding in nonrandomized longitudinal
  studies: a methodological review.
\newblock {\em Journal of clinical epidemiology}, 87:23--34.

\bibitem[Sussman and Ginsberg, 1998]{sussman1998rethinking}
Sussman, N. and Ginsberg, D. (1998).
\newblock Rethinking side effects of the selective serotonin reuptake
  inhibitors: sexual dysfunction and weight gain.
\newblock {\em Psychiatric Annals}, 28(2):89--97.

\bibitem[Zhu et~al., 2017]{zhu2017estimation}
Zhu, Y., Lawless, J.~F., and Cotton, C.~A. (2017).
\newblock Estimation of parametric failure time distributions based on
  interval-censored data with irregular dependent follow-up.
\newblock {\em Statistics in Medicine}, 36(10):1548--1567.

\end{thebibliography}

\newpage

\textbf{Supplementary Material for ``Accounting for Informative Monitoring Times under an Endogenous Covariate Process.''  }\vspace{0.2cm}\\

\newpage 
\noindent
\textbf{Supplementary Material A}
\textbf{Causal diagrams and biasing paths due to the monitoring process }\vspace{0.2cm}\\ 
In this section, we review the causal diagrams from the main manuscript, draw the biasing paths due to conditioning on the visit process, and show how models for the visit can be used to break the dependence between the monitoring and the outcome processes.

We first review the causal diagram that corresponds to Figure 1 in the main manuscript. We depict that diagram in Supplementary Figure 1, before intervening on it in any way. In Supplementary Figure 2, for the same causal diagram, we depict the biasing paths (in bold) due to the conditioning on the visit indicator, $dN(t)$ for $t\in \left\{0, 1, 2\right\}$, which acts as a collider.  

 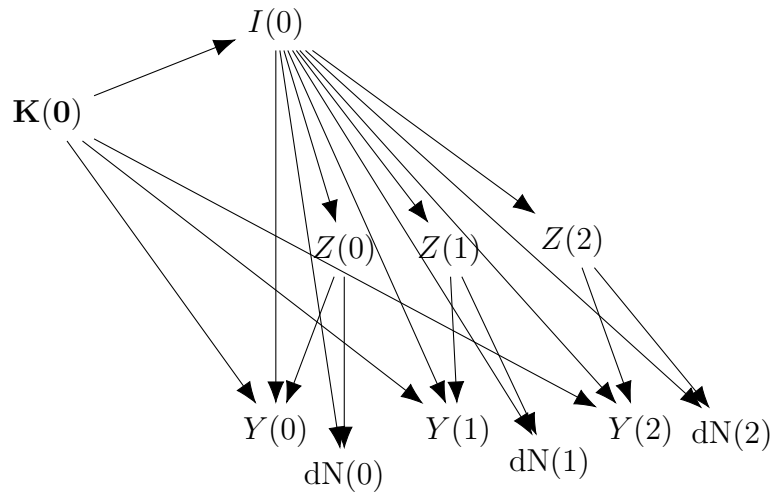
\begin{figure}[H]
\begin{center}
\begin{tikzpicture}[scale=0.6][%
->,
shorten >=2pt,
>=stealth,
node distance=1cm,
pil/.style={
->,
thick,
shorten =2pt,}
]


\node(1) at (-6,0){\textcolor{black}{$\mathbf{K(0)}$}};


\node (2) at (-1,2) {$ I(0) $};

 \node(3) at (-1,-7) {$Y(0)$};
 \node(3b) at (3,-7) {$Y(1)$};
 \node(3c) at (7,-7) {$Y(2)$};


\node(4) at (0.5, -3) {$Z(0)$};
\node(4b) at (2.8, -3) {$Z(1)$};
\node(4c) at (5.5, -2.8) {$Z(2)$};


\node(5) at (0.5,-8) {dN(0)};
\node(5b) at (5,-7.7) {dN(1)};
\node(5c) at (9,-7.1) {dN(2)};
 
\draw[-{Latex[length=3mm]}] (1) to (2);
\draw[-{Latex[length=3mm]}] (1) to (3);
   
\draw[-{Latex[length=3mm]}] (1) to (3b);
\draw[-{Latex[length=3mm]}] (1) to (3c);

\draw[-{Latex[length=3mm]}] (2) to (3);
\draw[-{Latex[length=3mm]}] (2) to (3b);
\draw[-{Latex[length=3mm]}] (2) to (3c);

\draw[-{Latex[length=3mm]}] (2) to (4);
 \draw[-{Latex[length=3mm]}] (4) to (3);

\draw[-{Latex[length=3mm]}] (2) to (4b);
 \draw[-{Latex[length=3mm]}] (4b) to (3b);

\draw[-{Latex[length=3mm]}] (2) to (4c);
 \draw[-{Latex[length=3mm]}] (4c) to (3c);

\draw[-{Latex[length=3mm]}] (2) to (5);
\draw[-{Latex[length=3mm]}] (2) to (5b);
\draw[-{Latex[length=3mm]}] (2) to (5c);
\draw[-{Latex[length=3mm]}] (4) to (5);
\draw[-{Latex[length=3mm]}] (4b) to (5b);
\draw[-{Latex[length=3mm]}] (4c) to (5c);

\end{tikzpicture}
\end{center}
\caption{Causal diagram for the first data generating mechanism (DGM) (patient index $i$ removed)} \label{fig1a}
\end{figure} 
 
 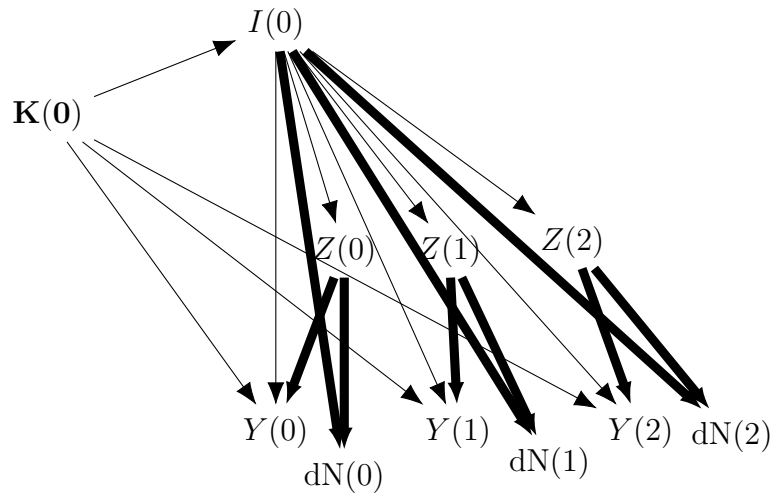
\begin{figure}[H]
\begin{center}
\begin{tikzpicture}[scale=0.6][%
->,
shorten >=2pt,
>=stealth,
node distance=1cm,
pil/.style={
->,
thick,
shorten =2pt,}
]


\node(1) at (-6,0){\textcolor{black}{$\mathbf{K(0)}$}};


\node (2) at (-1,2) {$ I(0) $};

 \node(3) at (-1,-7) {$Y(0)$};
 \node(3b) at (3,-7) {$Y(1)$};
 \node(3c) at (7,-7) {$Y(2)$};


\node(4) at (0.5, -3) {$Z(0)$};
\node(4b) at (2.8, -3) {$Z(1)$};
\node(4c) at (5.5, -2.8) {$Z(2)$};


\node(5) at (0.5,-8) {dN(0)};
\node(5b) at (5,-7.7) {dN(1)};
\node(5c) at (9,-7.1) {dN(2)};
 
\draw[-{Latex[length=3mm]}] (1) to (2);
\draw[-{Latex[length=3mm]}] (1) to (3);
   
\draw[-{Latex[length=3mm]}] (1) to (3b);
\draw[-{Latex[length=3mm]}] (1) to (3c);

\draw[-{Latex[length=3mm]}] (2) to (3);
\draw[-{Latex[length=3mm]}] (2) to (3b);
\draw[-{Latex[length=3mm]}] (2) to (3c);

\draw[-{Latex[length=3mm]}] (2) to (4);
 \draw[-{Latex[length=3mm]},line width=1.2mm] (4) to (3);

\draw[-{Latex[length=3mm]}] (2) to (4b);
 \draw[-{Latex[length=3mm]},line width=1.2mm] (4b) to (3b);

\draw[-{Latex[length=3mm]}] (2) to (4c);
 \draw[-{Latex[length=3mm]},line width=1.2mm] (4c) to (3c);

\draw[-{Latex[length=3mm]},line width=1.2mm] (2) to (5);
\draw[-{Latex[length=3mm]},line width=1.2mm] (2) to (5b);
\draw[-{Latex[length=3mm]},line width=1.2mm] (2) to (5c);
\draw[-{Latex[length=3mm]},line width=1.2mm] (4) to (5);
\draw[-{Latex[length=3mm]},line width=1.2mm] (4b) to (5b);
\draw[-{Latex[length=3mm]},line width=1.2mm] (4c) to (5c);

\end{tikzpicture}
\end{center}
\caption{Causal diagram for the first DGM (patient index $i$ removed), biaising paths in bold} \label{fig1b}
\end{figure} 
Finally, we depict in Supplementary Figure 3 what remains from the biasing paths (in bold) after adjusting for the monitoring rate via an inverse monitoring rate conditional on covariates $\mathbf{Z}$ and $\mathbf{I}$.   We find that there is no more unblocked path from the exposure to the outcome due to the monitoring process, that is not due to the marginal effect of treatment.   A proper adjustment for confounding factors $\mathbf{K(0)}$ must also be done to obtain unbiased estimates of the marginal effect of treatment (e.g. via inverse probability of treatment weights).
 
 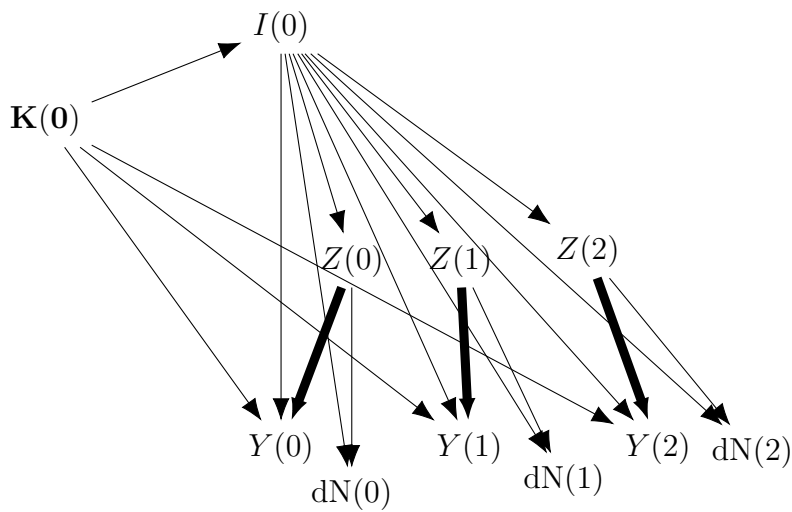
\begin{figure}[H]
\begin{center}
\begin{tikzpicture}[scale=0.62][%
->,
shorten >=2pt,
>=stealth,
node distance=1cm,
pil/.style={
->,
thick,
shorten =2pt,}
]


\node(1) at (-6,0){\textcolor{black}{$\mathbf{K(0)}$}};


\node (2) at (-1,2) {$ I(0) $};

 \node(3) at (-1,-7) {$Y(0)$};
 \node(3b) at (3,-7) {$Y(1)$};
 \node(3c) at (7,-7) {$Y(2)$};


\node(4) at (0.5, -3) {$Z(0)$};
\node(4b) at (2.8, -3) {$Z(1)$};
\node(4c) at (5.5, -2.8) {$Z(2)$};


\node(5) at (0.5,-8) {dN(0)};
\node(5b) at (5,-7.7) {dN(1)};
\node(5c) at (9,-7.1) {dN(2)};
 
\draw[-{Latex[length=3mm]}] (1) to (2);
\draw[-{Latex[length=3mm]}] (1) to (3);
   
\draw[-{Latex[length=3mm]}] (1) to (3b);
\draw[-{Latex[length=3mm]}] (1) to (3c);

\draw[-{Latex[length=3mm]}] (2) to (3);
\draw[-{Latex[length=3mm]}] (2) to (3b);
\draw[-{Latex[length=3mm]}] (2) to (3c);

\draw[-{Latex[length=3mm]}] (2) to (4);
 \draw[-{Latex[length=3mm]},line width=1.2mm] (4) to (3);

\draw[-{Latex[length=3mm]}] (2) to (4b);
 \draw[-{Latex[length=3mm]},line width=1.2mm] (4b) to (3b);

\draw[-{Latex[length=3mm]}] (2) to (4c);
 \draw[-{Latex[length=3mm]},line width=1.2mm] (4c) to (3c);

\draw[-{Latex[length=3mm]} ] (2) to (5);
\draw[-{Latex[length=3mm]} ] (2) to (5b);
\draw[-{Latex[length=3mm]} ] (2) to (5c);
\draw[-{Latex[length=3mm]} ] (4) to (5);
\draw[-{Latex[length=3mm]} ] (4b) to (5b);
\draw[-{Latex[length=3mm]} ] (4c) to (5c);

\end{tikzpicture}
\end{center}
\caption{Causal diagram for the first DGM (patient index $i$ removed) after adjusting for the visit process} \label{fig1c}
\end{figure} 
 
We now review the causal diagram that corresponds to Figure 2 in the main manuscript. That diagram is depicted in Supplementary Figure 4, before intervening on it in any way. In Supplementary Figure 5, for the same causal diagram, we depict the biasing paths (in bold) due to the conditioning on the visit indicator, $dN(t)$ for $t\in \left\{1, 2, 3\right\}$, which acts as a collider.  
 
 \begin{figure}[H]
\begin{center}
\begin{tikzpicture}[scale=0.73][%
 ->,
shorten >=2pt,
>=stealth,
node distance=1cm,
pil/.style={
->,
thick,
shorten =2pt,}
]

\node (1) at (-1.4,0.4) {$ I(0) $};
 
 \node(3b) at (5,-2) {$Y(1)$};
\node(3c) at (9.5,-2) {$Y(2)$};
\node(3d) at (14,-2) {$Y(3)$};
\node(5) at (2,1.4){\textcolor{black}{$\mathbf{K(0)}$}};

\draw[-{Latex[length=3mm]}] (1) to  (3b);
\draw[-{Latex[length=3mm]}] (1) to  (3c);
\draw[-{Latex[length=3mm]}](5) to  (1);
 
 \draw[-{Latex[length=3mm]}](5) to (3b);
\draw[-{Latex[length=3mm]}](5) to (3c);
 
\node(30) at (-1.5, -3.8) {$Z(0)$};
 
\node(307) at (9, -4.6) {$Z(2)$};

\draw[-{Latex[length=3mm]}](1) to (30);

 \draw[-{Latex[length=3mm]}] (307) to (3d);
 
\node(11) at (5,-6) {\textcolor{black}{$dN(1)$}};
 \node(12) at (8.5,-6.6) {\textcolor{black}{$dN(2)$}};
   \node(13) at (12,-7.7) {\textcolor{black}{$dN(3)$}};
 \draw[-{Latex[length=3mm]}] (307) to (13);
\draw[-{Latex[length=3mm]}](5) to (3d);

 \draw[-{Latex[length=3mm]}] (1) to (307);
\draw[-{Latex[length=3mm]}](1) to (11);
\draw[-{Latex[length=3mm]}](1) to (12);
\draw[-{Latex[length=3mm]} ](30) to (11);
\draw[-{Latex[length=3mm]}](1) to (3d);

\draw[-{Latex[length=3mm]}](30) to (3b);
 
 \draw[-{Latex[length=3mm]}](30) to (3c);
 \draw[-{Latex[length=3mm]}](30) to (12);

 \draw[-{Latex[length=3mm]}] (1) to (13);
 
\end{tikzpicture}
\end{center}
\caption{Causal diagram for the second DGM (patient index $i$ removed)} \label{fig2a}
\end{figure}
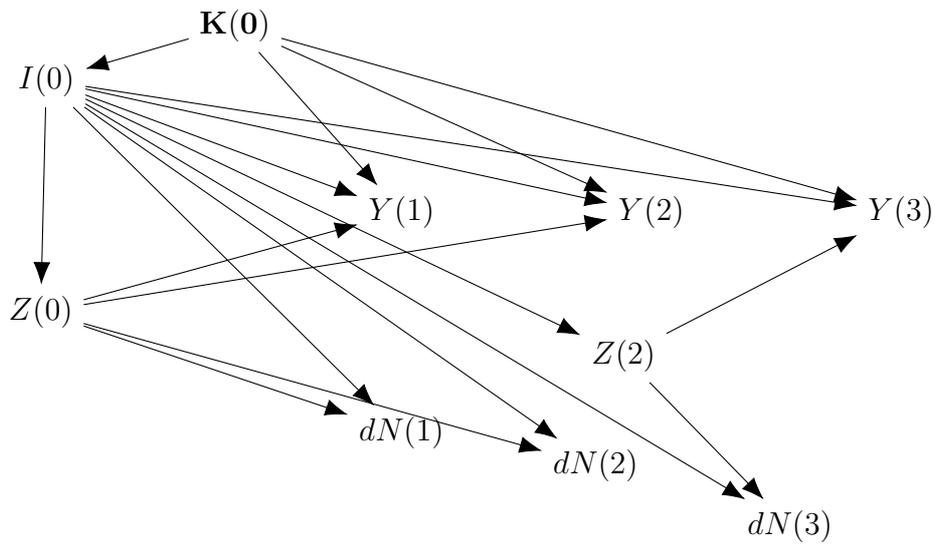 
We depict in Supplementary Figure 6 what remains from the biasing paths (in bold) after adjusting for the monitoring rate via an inverse monitoring rate conditional on covariates $\mathbf{Z}$ (for the last values of them, which we assume affect the monitoring indicator and the outcome) and $\mathbf{I}$.   We find that there is no more unblocked path from the exposure to the outcome due to the monitoring process, that is not due to the marginal effect of treatment. In that case too, a proper adjustment for confounding factors $\mathbf{K(0)}$ must be done to obtain unbiased estimates of the marginal effect of treatment (e.g. via inverse probability of treatment weights).
 
 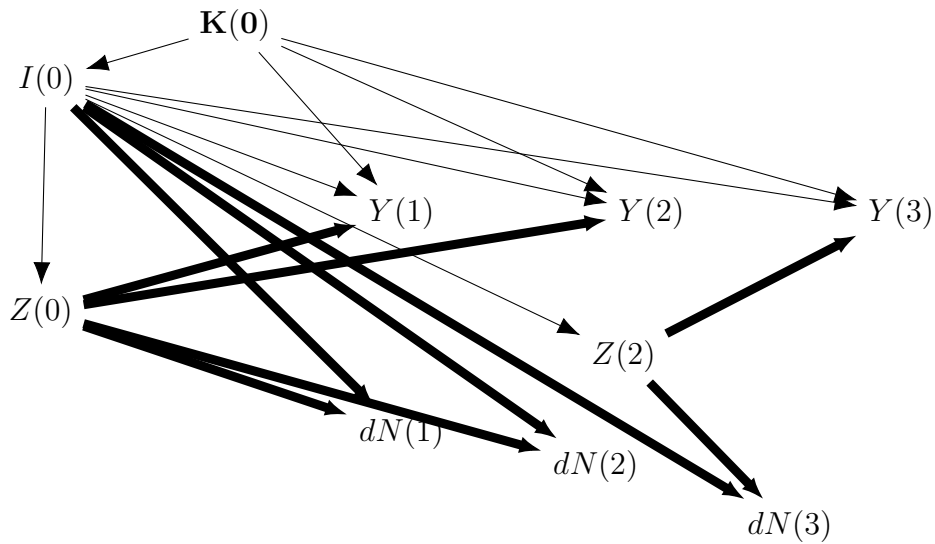
\begin{figure}[H]
\begin{center}
\begin{tikzpicture}[scale=0.73][%
  ->,
shorten >=2pt,
>=stealth,
node distance=1cm,
pil/.style={
->,
thick,
shorten =2pt,}
]

\node (1) at (-1.4,0.4) {$ I(0) $};
 
 \node(3b) at (5,-2) {$Y(1)$};
\node(3c) at (9.5,-2) {$Y(2)$};
\node(3d) at (14,-2) {$Y(3)$};
\node(5) at (2,1.4){\textcolor{black}{$\mathbf{K(0)}$}};

\draw[-{Latex[length=3mm]}] (1) to  (3b);
\draw[-{Latex[length=3mm]}] (1) to  (3c);
\draw[-{Latex[length=3mm]}](5) to  (1);
 
 \draw[-{Latex[length=3mm]}](5) to (3b);
\draw[-{Latex[length=3mm]}](5) to (3c);
 
\node(30) at (-1.5, -3.8) {$Z(0)$};
 
\node(307) at (9, -4.6) {$Z(2)$};

\draw[-{Latex[length=3mm]}](1) to (30);

 \draw[-{Latex[length=3mm]},line width=1.2mm] (307) to (3d);
 
\node(11) at (5,-6) {\textcolor{black}{$dN(1)$}};
 \node(12) at (8.5,-6.6) {\textcolor{black}{$dN(2)$}};
   \node(13) at (12,-7.7) {\textcolor{black}{$dN(3)$}};
 \draw[-{Latex[length=3mm]},line width=1.2mm] (307) to (13);
\draw[-{Latex[length=3mm]}](5) to (3d);

 \draw[-{Latex[length=3mm]}] (1) to (307);
\draw[-{Latex[length=3mm]},line width=1.2mm](1) to (11);
\draw[-{Latex[length=3mm]},line width=1.2mm](1) to (12);
\draw[-{Latex[length=3mm]} ,line width=1.2mm](30) to (11);
\draw[-{Latex[length=3mm]}](1) to (3d);

\draw[-{Latex[length=3mm]},line width=1.2mm](30) to (3b);
 
 \draw[-{Latex[length=3mm]},line width=1.2mm](30) to (3c);
 \draw[-{Latex[length=3mm]},line width=1.2mm](30) to (12);

 \draw[-{Latex[length=3mm]},line width=1.2mm] (1) to (13);

\end{tikzpicture}
\end{center}
\caption{Causal diagram for the second DGM (patient index $i$ removed), biaising paths in bold} \label{fig2b}
\end{figure} 
 
 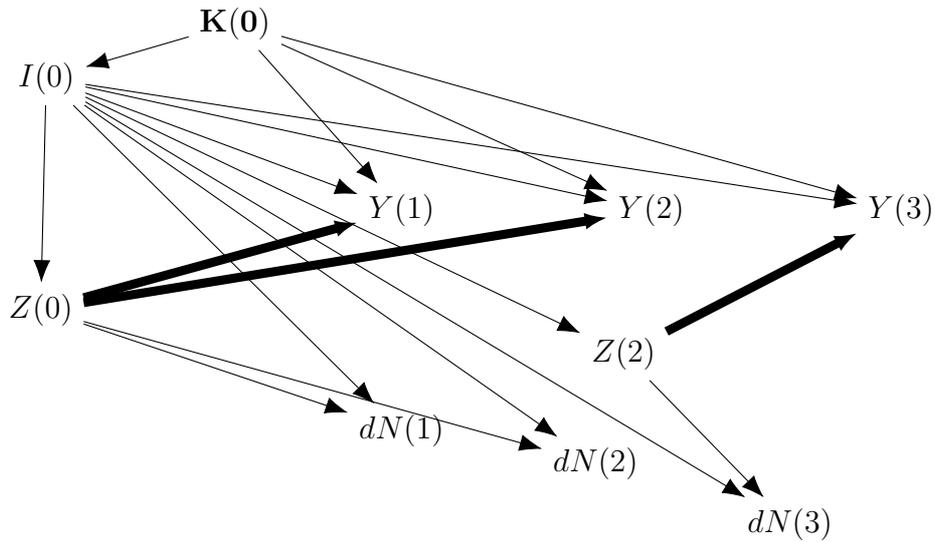
\begin{figure}[H]
\begin{center}
\begin{tikzpicture}[scale=0.73][%
  ->,
shorten >=2pt,
>=stealth,
node distance=1cm,
pil/.style={
->,
thick,
shorten =2pt,}
]

\node (1) at (-1.4,0.4) {$ I(0) $};
 
 \node(3b) at (5,-2) {$Y(1)$};
\node(3c) at (9.5,-2) {$Y(2)$};
\node(3d) at (14,-2) {$Y(3)$};
\node(5) at (2,1.4){\textcolor{black}{$\mathbf{K(0)}$}};

\draw[-{Latex[length=3mm]}] (1) to  (3b);
\draw[-{Latex[length=3mm]}] (1) to  (3c);
\draw[-{Latex[length=3mm]}](5) to  (1);
 
 \draw[-{Latex[length=3mm]}](5) to (3b);
\draw[-{Latex[length=3mm]}](5) to (3c);
 
\node(30) at (-1.5, -3.8) {$Z(0)$};
 
\node(307) at (9, -4.6) {$Z(2)$};

\draw[-{Latex[length=3mm]}](1) to (30);

 \draw[-{Latex[length=3mm]},line width=1.2mm] (307) to (3d);
 
\node(11) at (5,-6) {\textcolor{black}{$dN(1)$}};
 \node(12) at (8.5,-6.6) {\textcolor{black}{$dN(2)$}};
   \node(13) at (12,-7.7) {\textcolor{black}{$dN(3)$}};
 \draw[-{Latex[length=3mm]}] (307) to (13);
\draw[-{Latex[length=3mm]}](5) to (3d);

 \draw[-{Latex[length=3mm]}] (1) to (307);
\draw[-{Latex[length=3mm]} ](1) to (11);
\draw[-{Latex[length=3mm]} ](1) to (12);
\draw[-{Latex[length=3mm]}  ](30) to (11);
\draw[-{Latex[length=3mm]}](1) to (3d);

\draw[-{Latex[length=3mm]},line width=1.2mm](30) to (3b);
 
 \draw[-{Latex[length=3mm]},line width=1.2mm](30) to (3c);
 \draw[-{Latex[length=3mm]}](30) to (12);

 \draw[-{Latex[length=3mm]} ] (1) to (13);

\end{tikzpicture}
\end{center}
\caption{Causal diagram for the second DGM (patient index $i$ removed), remainings of the biaising paths in bold} \label{fig2c}
\end{figure}

Now, we review the causal diagram that corresponds to Figure 3 in the main manuscript. That diagram is depicted in Supplementary Figure 7, where we show the diagram before intervening on it. In Supplementary Figure 8, we depict for the same diagram the biasing paths (in bold) due to the conditioning on the visit indicator, $dN(t)$ for $t\in \left\{1, 2\right\}$, which acts as a collider (Note: in this document, dashed edges are used to make it clearer which distinct paths can bias the estimate of the marginal effect of treatment, but they do not bear any special or different meaning as compared to other paths or causal arrows).

 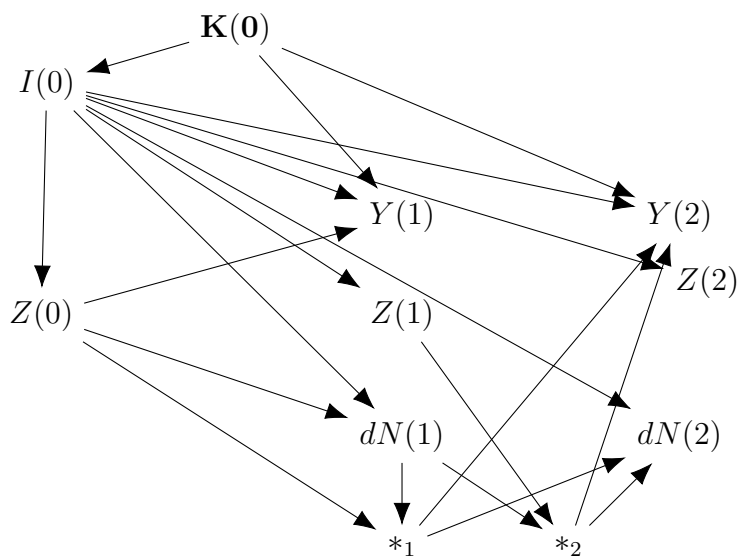
\begin{figure}[H]
\begin{center}
\begin{tikzpicture}[scale=0.73][%
->,
shorten >=2pt,
>=stealth,
node distance=1cm,
pil/.style={
->,
thick,
shorten =2pt,}
]

\node (1) at (-1.4,0.4) {$ I(0) $};
 
 \node(3b) at (5,-2) {$Y(1)$};
\node(3c) at (10,-2) {$Y(2)$};
 
\node(5) at (2,1.4){\textcolor{black}{$\mathbf{K(0)}$}};
 
\draw[-{Latex[length=3mm]}] (1) to  (3b);
\draw[-{Latex[length=3mm]}] (1) to  (3c);
\draw[-{Latex[length=3mm]},black](5) to  (1);
 
 \draw[-{Latex[length=3mm]},black](5) to (3b);
\draw[-{Latex[length=3mm]},black](5) to (3c);
 
\node(30) at (-1.5, -3.8) {$Z(0)$};
\node(31) at (5, -3.8) {$Z(1)$};
 \node(32) at (10.5, -3.2) {$Z(2)$};

\draw[-{Latex[length=3mm]}](1) to (30);
\draw[-{Latex[length=3mm]}](1) to (31);
 \draw[-{Latex[length=3mm]}](1) to (32);

\node(11) at (5,-6) {\textcolor{black}{$dN(1)$}};
 \node(12) at (10,-6) {\textcolor{black}{$dN(2)$}};
 
 \node(s1) at (5,-8) {$*_1$};
\draw[-{Latex[length=3mm]}](30)  to (s1);
 \draw[-{Latex[length=3mm]}](11)  to (s1);
 \node(s2) at (8,-8) {$*_2$};
\draw[-{Latex[length=3mm]}](31)  to (s2);
 \draw[-{Latex[length=3mm]}](11)  to (s2);
 
\draw[-{Latex[length=3mm]}](1) to (11);
\draw[-{Latex[length=3mm]}](1) to (12);
\draw[-{Latex[length=3mm]}](30) to (11);

\draw[-{Latex[length=3mm]}](30) to (3b);
 
\draw[-{Latex[length=3mm]}](s1) to (12);
 \draw[-{Latex[length=3mm]}](s2) to (12);

\draw[-{Latex[length=3mm]}](s1) to (3c);
 \draw[-{Latex[length=3mm]}](s2) to (3c);

\end{tikzpicture}
\end{center}
\caption{Causal diagram for the third DGM (patient index $i$ removed)} \label{fig3a}
\end{figure} 

 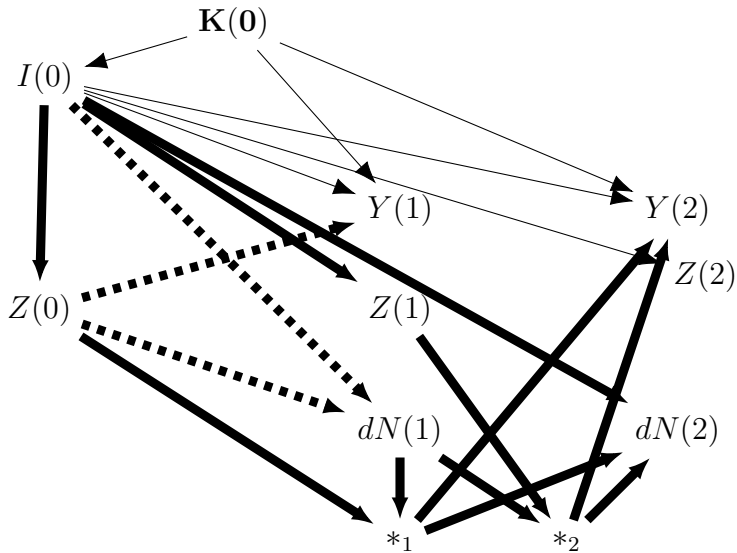
\begin{figure}[H]
\begin{center}
\begin{tikzpicture}[scale=0.73][%
->,
shorten >=2pt,
>=stealth,
node distance=1cm,
pil/.style={
->,
thick,
shorten =2pt,}
]

\node (1) at (-1.4,0.4) {$ I(0) $};
 
 \node(3b) at (5,-2) {$Y(1)$};
\node(3c) at (10,-2) {$Y(2)$};
 
\node(5) at (2,1.4){\textcolor{black}{$\mathbf{K(0)}$}};
 
\draw[-{Latex[length=3mm]}] (1) to  (3b);
\draw[-{Latex[length=3mm]}] (1) to  (3c);
\draw[-{Latex[length=3mm]},black](5) to  (1);
 
 \draw[-{Latex[length=3mm]},black](5) to (3b);
\draw[-{Latex[length=3mm]},black](5) to (3c);
 
\node(30) at (-1.5, -3.8) {$Z(0)$};
\node(31) at (5, -3.8) {$Z(1)$};
 \node(32) at (10.5, -3.2) {$Z(2)$};

\draw[,-{Latex[length=3mm]}, line width=1.2mm](1) to (30);
\draw[-{Latex[length=3mm]}, line width=1.2mm](1) to (31);
 \draw[-{Latex[length=3mm]}](1) to (32);

\node(11) at (5,-6) {\textcolor{black}{$dN(1)$}};
 \node(12) at (10,-6) {\textcolor{black}{$dN(2)$}};
  
 \node(s1) at (5,-8) {$*_1$};
\draw[-{Latex[length=3mm]}, line width=1.2mm](30)  to (s1);
 \draw[-{Latex[length=3mm]}, line width=1.2mm](11)  to (s1);
 \node(s2) at (8,-8) {$*_2$};
\draw[-{Latex[length=3mm]}, line width=1.2mm](31)  to (s2);
 \draw[-{Latex[length=3mm]}, line width=1.2mm](11)  to (s2);
 
\draw[dashed,-{Latex[length=3mm]}, line width=1.2mm](1) to (11);
\draw[-{Latex[length=3mm]}, line width=1.2mm](1) to (12);
\draw[dashed,-{Latex[length=3mm]}, line width=1.2mm](30) to (11);

\draw[dashed,-{Latex[length=3mm]}, line width=1.2mm](30) to (3b);
 
\draw[-{Latex[length=3mm]}, line width=1.2mm](s1) to (12);
 \draw[-{Latex[length=3mm]}, line width=1.2mm](s2) to (12);

\draw[-{Latex[length=3mm]}, line width=1.2mm](s1) to (3c);
 \draw[-{Latex[length=3mm]}, line width=1.2mm](s2) to (3c);
 
\end{tikzpicture}
\end{center}
\caption{Causal diagram for the third DGM (patient index $i$ removed), biaising paths in bold. Dashed edges are used to make it clearer which distinct paths can bias the estimate of the marginal effect of treatment} \label{fig3b}
\end{figure} 

We now make a dinstinction between two scenarios for the visit pattern, again assuming that we are in the setting depicted in Supplementary Figure 8: (a) There is a visit at time 1 and $dN(1)=1$, and (b) there is no visit at time 1 and $dN(1)=0$. In the former case (a), suppose we only adjust for an inverse intensity weight as a function of the last covariates observed. In Supplementary Figure 9, we depict what remains from the biasing paths after adjusting for the monitoring rate via an inverse monitoring rate conditional on the last covariates $\mathbf{Z}$ and $\mathbf{I}$ observed (in bold). We find that there is yet at least one unblocked path from the exposure to the outcome $Y(2)$ that is not due to the marginal effect of treatment and that is due to conditioning on collider $dN(2)$ (the path is given by $I(0) \rightarrow Z(0) \rightarrow \bm{*_1} - dN(2) - \bm{*_2} \rightarrow Y(2)$). 
 
  \begin{figure}[H]
\begin{center}
\begin{tikzpicture}[scale=0.73][%
->,
shorten >=2pt,
>=stealth,
node distance=1cm,
pil/.style={
->,
thick,
shorten =2pt,}
]

\node (1) at (-1.4,0.4) {$ I(0) $};
 
 \node(3b) at (5,-2) {$Y(1)$};
\node(3c) at (10,-2) {$Y(2)$};
 
\node(5) at (2,1.4){\textcolor{black}{$\mathbf{K(0)}$}};
 
\draw[-{Latex[length=3mm]}] (1) to  (3b);
\draw[-{Latex[length=3mm]}] (1) to  (3c);
\draw[-{Latex[length=3mm]},black](5) to  (1);
 
 \draw[-{Latex[length=3mm]},black](5) to (3b);
\draw[-{Latex[length=3mm]},black](5) to (3c);
 
\node(30) at (-1.5, -3.8) {$Z(0)$};
\node(31) at (5, -3.8) {$Z(1)$};
 \node(32) at (10.5, -3.2) {$Z(2)$};

\draw[,-{Latex[length=3mm]}, line width=1.2mm](1) to (30);
\draw[-{Latex[length=3mm]}, line width=1.2mm](1) to (31);
 \draw[-{Latex[length=3mm]}](1) to (32);

\node(11) at (5,-6) {\textcolor{black}{$dN(1)$}};
 \node(12) at (10,-6) {\textcolor{black}{$dN(2)$}};
  
 \node(s1) at (5,-8) {$*_1$};
\draw[-{Latex[length=3mm]}, line width=1.2mm](30)  to (s1);
 \draw[-{Latex[length=3mm]}, line width=1.2mm](11)  to (s1);
 \node(s2) at (8,-8) {$*_2$};
\draw[-{Latex[length=3mm]} ](31)  to (s2);
 \draw[-{Latex[length=3mm]}, line width=1.2mm](11)  to (s2);
 
\draw[dashed,-{Latex[length=3mm]}](1) to (11);
\draw[-{Latex[length=3mm]}](1) to (12);
\draw[dashed,-{Latex[length=3mm]}](30) to (11);

\draw[dashed,-{Latex[length=3mm]}, line width=1.2mm](30) to (3b);
 
\draw[-{Latex[length=3mm]}, line width=1.2mm](s1) to (12);
 \draw[-{Latex[length=3mm]}, line width=1.2mm](s2) to (12);

\draw[-{Latex[length=3mm]}](s1) to (3c);
 \draw[-{Latex[length=3mm]}, line width=1.2mm](s2) to (3c);

\end{tikzpicture}
\end{center}
 \caption{Causal diagram for the third DGM (patient index $i$ removed), remainings of the biaising paths in bold in scenario (a) when adjusting for the visit process using an inverse intensity weight as a function of the last covariates observed} \label{fig3c1}
\end{figure}
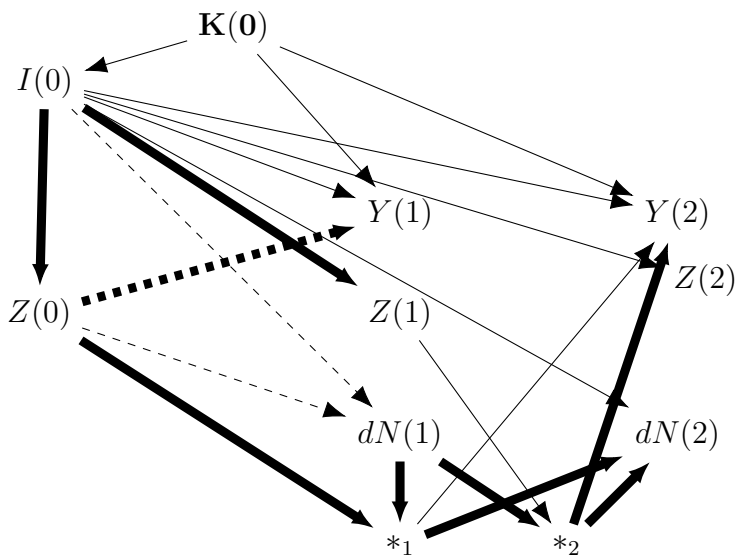 
In scenario (b), where there is no visit at time 1, suppose we only adjust for an inverse intensity weight as a function of the last covariates observed. What remains from the biasing paths after adjusting for the monitoring rate via an inverse monitoring rate conditional on the last covariates $\mathbf{Z}$ and $\mathbf{I}$ observed is depicted in Supplementary Figure 10. Here again, there is yet an unblocked path from the exposure to the outcome $Y(2)$ that is not due to the marginal effect of treatment (given by $I(0) \rightarrow Z(1) \rightarrow \bm{*_2} - dN(2) - \bm{*_1} \rightarrow Y(2)$).
 
    Now, suppose that we use a cumulated weight that accounts for the full history of covariates and their interaction with the monitoring process (such as the weight $sw_{i,j}(\cdot)$ proposed in the manuscript). The Supplemetary Figure 11 depicts, for scenario (a), the remaining parts of the biasing paths in bold in such case; the adjustment for the whole monitoring process effectively blocks the unblocked path between subsequent monitoring indicators, which paths were ``due'' to the interaction terms. We depicted this by removing the bold from the arrows leaving interaction terms $\bm{*_1}$ and $\bm{*_2}$, and entering the node $dN(2)$ but this could probably be depicted differently too (e.g. by removing the whole path between $dN(1)$ and $dN(2)$). The former unblocked path is now blocked by e.g. the interaction term $*_1$ which is a collider. 
  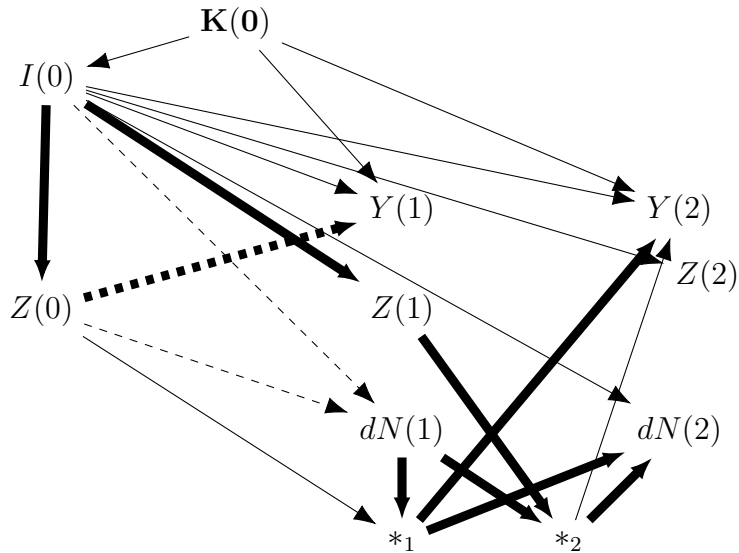
\begin{figure}[H]
\begin{center}
\begin{tikzpicture}[scale=0.73][%
 ->,
shorten >=2pt,
>=stealth,
node distance=1cm,
pil/.style={
->,
thick,
shorten =2pt,}
]

\node (1) at (-1.4,0.4) {$ I(0) $};
 
 \node(3b) at (5,-2) {$Y(1)$};
\node(3c) at (10,-2) {$Y(2)$};
 
\node(5) at (2,1.4){\textcolor{black}{$\mathbf{K(0)}$}};
 
\draw[-{Latex[length=3mm]}] (1) to  (3b);
\draw[-{Latex[length=3mm]}] (1) to  (3c);
\draw[-{Latex[length=3mm]},black](5) to  (1);
 
 \draw[-{Latex[length=3mm]},black](5) to (3b);
\draw[-{Latex[length=3mm]},black](5) to (3c);
 
\node(30) at (-1.5, -3.8) {$Z(0)$};
\node(31) at (5, -3.8) {$Z(1)$};
 \node(32) at (10.5, -3.2) {$Z(2)$};

\draw[,-{Latex[length=3mm]}, line width=1.2mm](1) to (30);
\draw[-{Latex[length=3mm]}, line width=1.2mm](1) to (31);
 \draw[-{Latex[length=3mm]}](1) to (32);

\node(11) at (5,-6) {\textcolor{black}{$dN(1)$}};
 \node(12) at (10,-6) {\textcolor{black}{$dN(2)$}};
  
 \node(s1) at (5,-8) {$*_1$};
\draw[-{Latex[length=3mm]}](30)  to (s1);
 \draw[-{Latex[length=3mm]}, line width=1.2mm](11)  to (s1);
 \node(s2) at (8,-8) {$*_2$};
\draw[-{Latex[length=3mm]}, line width=1.2mm](31)  to (s2);
 \draw[-{Latex[length=3mm]}, line width=1.2mm](11)  to (s2);
 
\draw[dashed,-{Latex[length=3mm]}](1) to (11);
\draw[-{Latex[length=3mm]}](1) to (12);
\draw[dashed,-{Latex[length=3mm]}](30) to (11);

\draw[dashed,-{Latex[length=3mm]}, line width=1.2mm](30) to (3b);
 
\draw[-{Latex[length=3mm]}, line width=1.2mm](s1) to (12);
 \draw[-{Latex[length=3mm]}, line width=1.2mm](s2) to (12);

\draw[-{Latex[length=3mm]}, line width=1.2mm](s1) to (3c);
 \draw[-{Latex[length=3mm]}](s2) to (3c);

\end{tikzpicture}
\end{center}
 \caption{Causal diagram for the third DGM (patient index $i$ removed), remainings of the biaising path in bold in scenario (b) when adjusting for the visit process using an inverse intensity weight as a function of the last covariates observed } \label{fig3c2}
\end{figure} 
 For scenario (b), Supplementary Figure 12 depicts the remaining biasing paths after using the proposed cumulated weight. There again, we depicted the impact of adjusting for the whole monitoring process by removing the bold from the arrows leaving interaction terms $\bm{*_1}$ and $\bm{*_2}$, and entering the node $dN(2)$; the formerly unblocked biasing path is now blocked by e.g. the collider $*_2$. In the two figures (Supplementary Figures 11 and 12), there is no more biasing path from the exposure to the outcome that would be due to colliders $dN(\cdot)$ after using the proposed cumulated weight.
   \begin{figure}[H]
\begin{center}
\begin{tikzpicture}[scale=0.7][%
->,
shorten >=2pt,
>=stealth,
node distance=1cm,
pil/.style={
->,
thick,
shorten =2pt,}
]

\node (1) at (-1.4,0.4) {$ I(0) $};
 
 \node(3b) at (5,-2) {$Y(1)$};
\node(3c) at (10,-2) {$Y(2)$};
 
\node(5) at (2,1.4){\textcolor{black}{$\mathbf{K(0)}$}};
 
\draw[-{Latex[length=3mm]}] (1) to  (3b);
\draw[-{Latex[length=3mm]}] (1) to  (3c);
\draw[-{Latex[length=3mm]},black](5) to  (1);
 
 \draw[-{Latex[length=3mm]},black](5) to (3b);
\draw[-{Latex[length=3mm]},black](5) to (3c);
 
\node(30) at (-1.5, -3.8) {$Z(0)$};
\node(31) at (5, -3.8) {$Z(1)$};
 \node(32) at (10.5, -3.2) {$Z(2)$};

\draw[,-{Latex[length=3mm]}, line width=1.2mm](1) to (30);
\draw[-{Latex[length=3mm]}, line width=1.2mm](1) to (31);
 \draw[-{Latex[length=3mm]}](1) to (32);

\node(11) at (5,-6) {\textcolor{black}{$dN(1)$}};
 \node(12) at (10,-6) {\textcolor{black}{$dN(2)$}};
  
 \node(s1) at (5,-8) {$*_1$};
\draw[-{Latex[length=3mm]}, line width=1.2mm](30)  to (s1);
 \draw[-{Latex[length=3mm]}, line width=1.2mm](11)  to (s1);
 \node(s2) at (8,-8) {$*_2$};
\draw[-{Latex[length=3mm]} ](31)  to (s2);
 \draw[-{Latex[length=3mm]}, line width=1.2mm](11)  to (s2);
 
\draw[dashed,-{Latex[length=3mm]}](1) to (11);
\draw[-{Latex[length=3mm]}](1) to (12);
\draw[dashed,-{Latex[length=3mm]}](30) to (11);

\draw[dashed,-{Latex[length=3mm]}, line width=1.2mm](30) to (3b);
 
\draw[-{Latex[length=3mm]} ](s1) to (12);
 \draw[-{Latex[length=3mm]}](s2) to (12);

\draw[-{Latex[length=3mm]}](s1) to (3c);
 \draw[-{Latex[length=3mm]}, line width=1.2mm](s2) to (3c);

\end{tikzpicture}
\end{center}
 \caption{Causal diagram for the third DGM (patient index $i$ removed), remainings of the biaising paths in bold in scenario (a) when using the proposed cumulated weight } \label{fig3c3}
\end{figure}
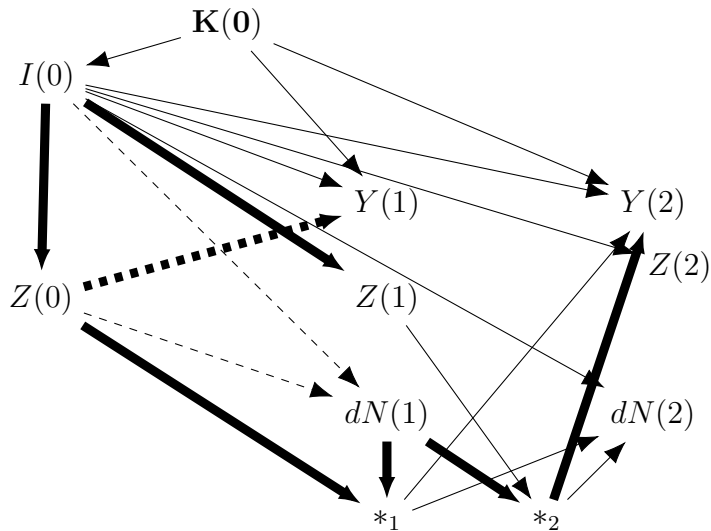 
 
  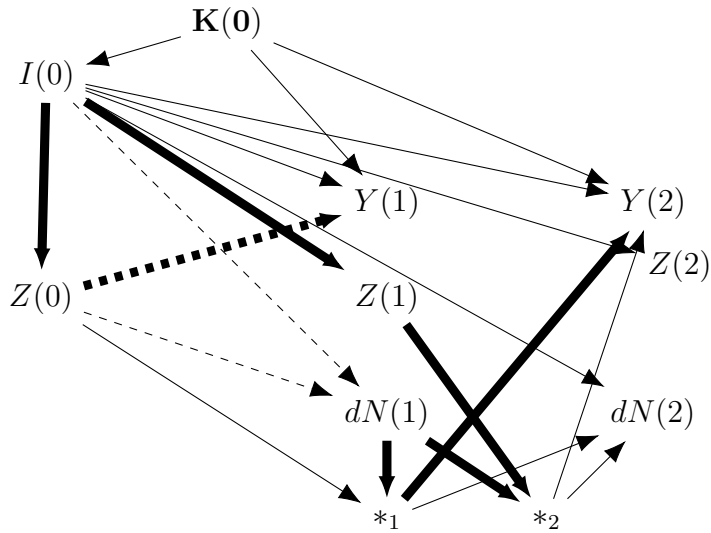
\begin{figure}[H]
\begin{center}
\begin{tikzpicture}[scale=0.7][%
 ->,
shorten >=2pt,
>=stealth,
node distance=1cm,
pil/.style={
->,
thick,
shorten =2pt,}
]

\node (1) at (-1.4,0.4) {$ I(0) $};
 
 \node(3b) at (5,-2) {$Y(1)$};
\node(3c) at (10,-2) {$Y(2)$};
 
\node(5) at (2,1.4){\textcolor{black}{$\mathbf{K(0)}$}};
 
\draw[-{Latex[length=3mm]}] (1) to  (3b);
\draw[-{Latex[length=3mm]}] (1) to  (3c);
\draw[-{Latex[length=3mm]},black](5) to  (1);
 
 \draw[-{Latex[length=3mm]},black](5) to (3b);
\draw[-{Latex[length=3mm]},black](5) to (3c);
 
\node(30) at (-1.5, -3.8) {$Z(0)$};
\node(31) at (5, -3.8) {$Z(1)$};
 \node(32) at (10.5, -3.2) {$Z(2)$};

\draw[,-{Latex[length=3mm]}, line width=1.2mm](1) to (30);
\draw[-{Latex[length=3mm]}, line width=1.2mm](1) to (31);
 \draw[-{Latex[length=3mm]}](1) to (32);

\node(11) at (5,-6) {\textcolor{black}{$dN(1)$}};
 \node(12) at (10,-6) {\textcolor{black}{$dN(2)$}};
  
 \node(s1) at (5,-8) {$*_1$};
\draw[-{Latex[length=3mm]}](30)  to (s1);
 \draw[-{Latex[length=3mm]}, line width=1.2mm](11)  to (s1);
 \node(s2) at (8,-8) {$*_2$};
\draw[-{Latex[length=3mm]}, line width=1.2mm](31)  to (s2);
 \draw[-{Latex[length=3mm]}, line width=1.2mm](11)  to (s2);
 
\draw[dashed,-{Latex[length=3mm]}](1) to (11);
\draw[-{Latex[length=3mm]}](1) to (12);
\draw[dashed,-{Latex[length=3mm]}](30) to (11);

\draw[dashed,-{Latex[length=3mm]}, line width=1.2mm](30) to (3b);
 
\draw[-{Latex[length=3mm]} ](s1) to (12);
 \draw[-{Latex[length=3mm]}](s2) to (12);

\draw[-{Latex[length=3mm]}, line width=1.2mm](s1) to (3c);
 \draw[-{Latex[length=3mm]}](s2) to (3c);

\end{tikzpicture}
\end{center}
 \caption{Causal diagram for the third DGM (patient index $i$ removed), remainings of the biaising paths in bold  in scenario (b) when using the proposed cumulated weight} \label{fig3c4}
\end{figure} 
  Finally, we present the last scenario, corresponding to Figure 4 in the main manuscript, which is similar to that depicted in Supplementary Figure 8, but where a previous outcome affects the next outcome and monitoring time. In Web Supplementary 13, we show the causal diagram corresponding to that scenario, before intervening on it. In Supplementary Figure 14, we depict the (potential) biasing paths (in bold) due to the conditioning on colliders dN(t), $t\subset 1, 2$. That setting (Supplementary Figures 13 and 14) is similar to that from Supplementary Figure 7, except for the collider $dN(2)$ that opens another path between $I(0)$ and $Y(1)$. By including the outcome as a predictor in the intensity model, a similar adjustment using the proposed cumulated weight with the intensity modelled conditionnally on $\mathbf{I}$, $\mathbf{Z}$ and $\mathbf{Y}$ will adjust properly for the visit process (not shown).
  \begin{figure}[H]
\begin{center}
\begin{tikzpicture}[scale=0.68][%
 ->,
shorten >=2pt,
>=stealth,
node distance=1cm,
pil/.style={
->,
thick,
shorten =2pt,}
]

\node (1) at (-1.4,0.4) {$ I(0) $};
 
 \node(3b) at (5,-2) {$Y(1)$};
\node(3c) at (10,-2) {$Y(2)$};
 
\node(5) at (2,1.4){\textcolor{black}{$\mathbf{K(0)}$}};

\draw[-{Latex[length=3mm]}] (1) to  (3b);
\draw[-{Latex[length=3mm]}] (1) to  (3c);
\draw[-{Latex[length=3mm]}](5) to  (1);
 
 \draw[-{Latex[length=3mm]}](5) to (3b);
\draw[-{Latex[length=3mm]}](5) to (3c);
 
\node(30) at (-1.5, -3.8) {$Z(0)$};
\node(31) at (5, -3.8) {$Z(1)$};
 \node(32) at (10.5, -3.2) {$Z(2)$};

\draw[-{Latex[length=3mm]}](1) to (30);
\draw[-{Latex[length=3mm]}](1) to (31);
 \draw[-{Latex[length=3mm]}](1) to (32);

\node(11) at (5,-6) {\textcolor{black}{$dN(1)$}};
 \node(12) at (10,-6) {\textcolor{black}{$dN(2)$}};

 \node(s1) at (5,-8) {$*_1$};
\draw[-{Latex[length=3mm]}](30)  to (s1);
 \draw[-{Latex[length=3mm]}](11)  to (s1);
 \node(s2) at (8,-8) {$*_2$};
\draw[-{Latex[length=3mm]}](31)  to (s2);
 \draw[-{Latex[length=3mm]}](11)  to (s2);

\draw[-{Latex[length=3mm]}](1) to (11);
\draw[-{Latex[length=3mm]}](1) to (12);
\draw[-{Latex[length=3mm]}](30) to (11);

\draw[-{Latex[length=3mm]}](30) to (3b);
 
\draw[-{Latex[length=3mm]}](s1) to (12);
 \draw[-{Latex[length=3mm]}](s2) to (12);

\draw[-{Latex[length=3mm]}](s1) to (3c);
 \draw[-{Latex[length=3mm]}](s2) to (3c);
 \draw[dashed, -{Latex[length=3mm]}](3b) to (3c);
 \draw[dashed, -{Latex[length=3mm]}](3b) to (12);
 
\end{tikzpicture}
\end{center}
 \caption{Causal diagram for the fourth DGM (patient index $i$ removed)} \label{fig4a}
\end{figure}
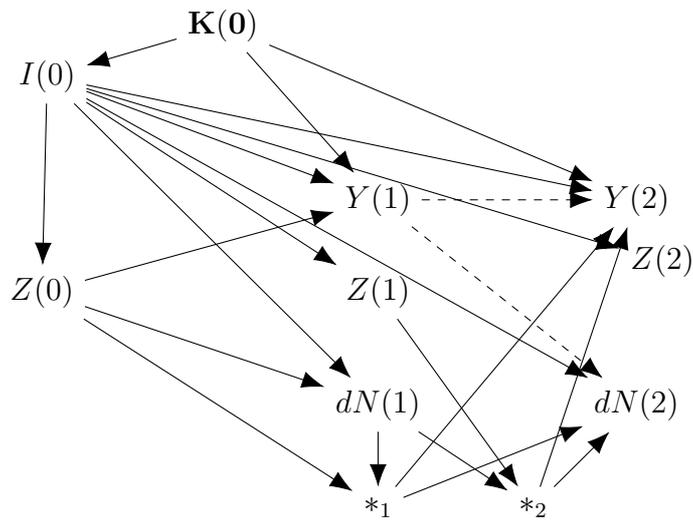

  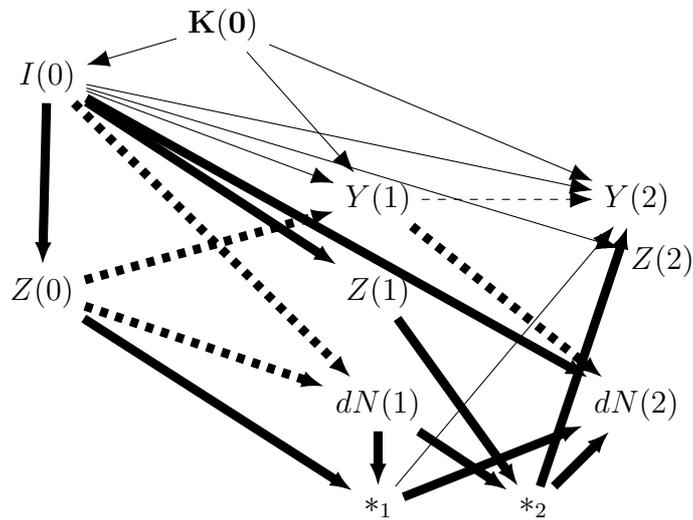
\begin{figure}[H]
\begin{center}
\begin{tikzpicture}[scale=0.68][%
 ->,
shorten >=2pt,
>=stealth,
node distance=1cm,
pil/.style={
->,
thick,
shorten =2pt,}
]

\node (1) at (-1.4,0.4) {$ I(0) $};
 
 \node(3b) at (5,-2) {$Y(1)$};
\node(3c) at (10,-2) {$Y(2)$};
 
\node(5) at (2,1.4){\textcolor{black}{$\mathbf{K(0)}$}};

\draw[-{Latex[length=3mm]}] (1) to  (3b);
\draw[-{Latex[length=3mm]}] (1) to  (3c);
\draw[-{Latex[length=3mm]}](5) to  (1);
 
 \draw[-{Latex[length=3mm]}](5) to (3b);
\draw[-{Latex[length=3mm]}](5) to (3c);
 
\node(30) at (-1.5, -3.8) {$Z(0)$};
\node(31) at (5, -3.8) {$Z(1)$};
 \node(32) at (10.5, -3.2) {$Z(2)$};

\draw[-{Latex[length=3mm]}, line width=1.2mm](1) to (30);
\draw[-{Latex[length=3mm]},line width=1.2mm](1) to (31);
 \draw[-{Latex[length=3mm]}](1) to (32);

\node(11) at (5,-6) {\textcolor{black}{$dN(1)$}};
 \node(12) at (10,-6) {\textcolor{black}{$dN(2)$}};

 \node(s1) at (5,-8) {$*_1$};
\draw[-{Latex[length=3mm]}, line width=1.2mm](30)  to (s1);
 \draw[-{Latex[length=3mm]},line width=1.2mm](11)  to (s1);
 \node(s2) at (8,-8) {$*_2$};
\draw[-{Latex[length=3mm]},line width=1.2mm](31)  to (s2);
 \draw[-{Latex[length=3mm]},line width=1.2mm](11)  to (s2);

\draw[-{Latex[length=3mm]}, line width=1.2mm, dashed](1) to (11);
\draw[-{Latex[length=3mm]},line width=1.2mm](1) to (12);
\draw[-{Latex[length=3mm]}, line width=1.2mm, dashed](30) to (11);

\draw[-{Latex[length=3mm]}, line width=1.2mm, dashed](30) to (3b);
 
\draw[-{Latex[length=3mm]},line width=1.2mm](s1) to (12);
 \draw[-{Latex[length=3mm]},line width=1.2mm](s2) to (12);

\draw[-{Latex[length=3mm]}](s1) to (3c);
 \draw[-{Latex[length=3mm]},line width=1.2mm](s2) to (3c);
 \draw[dashed, -{Latex[length=3mm]}](3b) to (3c);
 \draw[dashed, -{Latex[length=3mm]},line width=1.2mm](3b) to (12);
 
\end{tikzpicture}
\end{center}
 \caption{Causal diagram for the fourth DGM (patient index $i$ removed), biaising paths in bold} \label{fig4b}
\end{figure}

 \newpage
 
\textbf{Supplementary Material B}
\textbf{Estimating equation for the marginal effect of treatment on a continuous longitudinal outcome }\vspace{0.2cm}\\

\noindent In the main manuscript we assumed
\begin{align}
\tag{I3}I_i(0) &\perp \left\{ Y_{i0}(t), Y_{i1}(t)\right\} |  \mathbf{\mathcal{H}^o_i(t-)} , dN_i(t), \mathbf{K_i(0)},\\
\tag{I4}dN_i(t) &\perp Y_i(t) | \mathbf{\mathcal{H}^o_i(t-)}, \hspace{0.2cm}\text{and}  \\
\tag{I5}dN_i(t) &\perp dN_i(t-) | \mathbf{Z_i(l_i(t))}, I_i(0), B_i(t-), 
\end{align}
and proposed to use the following \textit{partial} model for the monitoring intensity:
\begin{align}
\tag{5} \lambda_i(t | \mathbf{Z_i(l(t_i))},  I_i(0), B_i(t-)) = &  \lambda_0(B_i(t)) \exp(\gamma_I I_i(0) + \bm{\gamma_Z }\mathbf{Z_i(l_i(t))}).
\end{align}
We also cited Theorem 1 of Pearl (2009), which we recall:

\noindent \textbf{Theorem 1} (The Causal Markov Condition). \textit{Any distribution generated by a Markovian model M can be factorized as:}
$$P(v_1, v_2, ..., v_n) = \prod_i P(v_i | pa_i)$$
\textit{ where }$V_1, V_2, ... V_n$ \textit{are the endogenous variables in M, and} $pa_i$ \textit{are (values of) the endogenous ``parents'' of $V_i$ in the causal diagram associated with M.}\vspace{0.1cm}\\
 
\noindent We assume that continuous time can be discretized in units of length 1 (e.g.~days) over which a visit can or cannot occur, so $t \in 1, 2, 3, ..., \lfloor \tau \rfloor$. We denote by $\overline{P(t)}$ the history from time 0 to time t of the covariate process $P$, and by $P(t) \setminus J(t)$ the process $P(t)$ minus the $J(t)$ process. Using the assumptions above, we have:
\begin{align}
\mathbb{E}\left[dN_i(t)  | \mathbf{\mathcal{H}^o_i(t-)} \right] =& \mathbb{P} ( dN_i(t)  |  \mathbf{\mathcal{H}^o_i(t-)} ) \nonumber \\
=& \mathbb{P} ( dN_i(t)  |  \mathbf{\mathcal{H}^o_i(t-)}\setminus \overline{dN_i(t-)}, \overline{dN_i(t-)} ) \nonumber \\
=&\frac{\mathbb{P} ( dN_i(t), \overline{dN_i(t-)}  |  \mathbf{\mathcal{H}^o_i(t-)}\setminus \overline{dN_i(t-)} )P( \mathbf{\mathcal{H}^o_i(t-)}\setminus \overline{dN_i(t-)} ) }{P( \mathbf{\mathcal{H}^o_i(t-)} )} \nonumber \\
\propto &\mathbb{P} ( \overline{dN_i(t)}   |  \mathbf{\mathcal{H}^o_i(t-)}\setminus \overline{dN_i(t-)} )\nonumber \\
=&\mathbb{P}(dN_i(t) |\mathbf{Z_i(l_i(t))}, I_i(0), B_i(t-1))  \nonumber\\
	& \hspace{0.4cm}\times \mathbb{P}(dN_i(t-1)|\mathbf{Z_i(l_i(t-1))}, I_i(0), B_i(t-2)) \nonumber \\
	&\hspace{0.8cm}	 \times ... \nonumber \\
		 & \hspace{1.1cm}		\times  \mathbb{P}(dN_i(2)| \mathbf{Z_i((l_i(1))}, I_i(0), B_i(1))   \nonumber\\
		 		&\hspace{1.4cm}	\times \mathbb{P}(dN_i(1)| \mathbf{Z_i(0)}, I_i(0))\nonumber\hspace{0.5cm} \text{using assumption (I5) and Theorem 1} \\
		 	=& \PRODI_{s=0}^t {\left\{ \xi_i(s) \exp \left( \gamma_I I_i(0) + \gamma_Z \mathbf{Z_i(l_i(s))}\right) \lambda_0(B_i(s)) ds \right\} }^{\mathbb{I}(dN_i(s)=1) }\nonumber \\
&\hspace{0.5cm}\times {\left\{1-  \xi_i(s) \exp \left( \gamma_I I_i(0) + \gamma_Z \mathbf{Z_i(l_i(s))}\right) \lambda_0(B_i(s)) ds  \right\} }^{\mathbb{I}(dN_i(s)=0)},
\end{align}
with the last term equal to the weight $usw_i(t|\mathbf{\mathcal{H}_i^o(t-) })$, and using assumption (5) for the visit model. As in the main manuscript, we use a product integral to emphasize that the product must be taken over continuous time. We use the following estimating equation for the coefficients $\bm{\beta_S}$, with the marginal effect of treatment consisting in e.g.~the last coefficient of $\bm{\beta_S}$:
\begin{align}
\mathbb{E} \left( \int_0^{\tau}\frac{ \mathbf{Y(t)} -  [\bm{\beta_s}' \mathbf{S(t)}]}{\mathbf{w(t|K(t))} \mathbf{usw}\mathbf{(t| \mathcal{H}^o(t-))}  } \mathbf{dN(t)} \right) =\mathbf{0}. \nonumber
\end{align} 
The matrix $\mathbf{S(t)}$ may, for instance, incorporate a column of 1 for estimating a constant intercept, or several columns as the basis of a cubic spline for accounting for the effect of time.

\noindent Using iterated expectation (similarly to Lin et al., 2004), we have that

 \begin{align}
&\mathbb{E} \left( \int_0^{\tau}\frac{ \mathbf{Y(t}) -  [\bm{\beta_s}' \mathbf{S(t)}]}{\mathbf{w(t|K(t))} \mathbf{usw}\mathbf{(t| \mathcal{H}^o(t-))}  } \mathbf{dN(t) }\right)\nonumber  \\
&\hspace{3cm} =\mathbb{E} \left[  \mathbb{E} \left( \int_0^{\tau}\frac{ \mathbf{Y(t}) -  [\bm{\beta_s}' \mathbf{S(t)}]}{\mathbf{w(t|K(t))} \mathbf{usw}\mathbf{(t| \mathcal{H}^o(t-))}  } \mathbf{dN(t}) \right) \mid \mathbf{\mathcal{H}^o(t-)}  \right]\nonumber \\
&\hspace{3cm} = \mathbb{E} \left( \int_0^{\tau}\frac{\mathbf{Y(t)} -  [\bm{\beta_s}' \mathbf{S(t)}]}{\mathbf{w(t|K(t))} \mathbf{usw(t| \mathcal{H}^o(t-))}  } \mathbb{E}\left[ \mathbf{dN(t)}| \mathbf{\mathcal{H}^o(t-)} \right] \right) \nonumber \text{using assumption (I4)}\\
&\hspace{3cm} \propto \mathbb{E}  \left( \int_0^{\tau}\frac{\mathbf{Y(t)} -  [\bm{\beta_s}' \mathbf{S(t)}]}{\mathbf{w(t|K(t))} \mathbf{usw(t| \mathcal{H}^o(t-))}  } \mathbf{usw(t|  \mathcal{H}^o(t-))}  dt  \right) \nonumber \\
&\hspace{3cm} = \mathbb{E}  \left( \int_0^{\tau}\frac{ \mathbf{Y(t)} -  [\bm{\beta_s}' \mathbf{S(t)}]}{\mathbf{w(t| K(t))} }dt  \right).\nonumber 
\end{align} 
Under assumption (I3) and correct model specifications, we have that the last expression in the final line is equal to 0, that is,
$$ \mathbb{E}  \left( \int_0^{\tau}\frac{ \mathbf{Y(t)} -  [\bm{\beta_s}' \mathbf{S(t)}]}{\mathbf{w(t|K(t))} }dt  \right) = \mathbf{0}$$
and the estimating equation for the marginal effect of treatment is unbiased.

\newpage

\textbf{Supplementary Material C}
\textbf{Asymptotic properties of the proposed estimator} \vspace{0.2cm}\\

\noindent Under the assumptions on the exposure and the monitoring models that follow:
\begin{align}
\tag{I3} I_i(t) & \perp \left\{ Y_{i0}(t), Y_{i1}(t)\right\} |  \mathbf{\mathcal{H}^o_i(t-)} , dN_i(t), \mathbf{K_i(t)} \label{v2}\\
\tag{I4} dN_i(t) &\perp Y_i(t) | \mathbf{\mathcal{H}^o_i(t-)}\\
\tag{I5} dN_i(t) &\perp dN_i(t-) | \mathbf{Z_i(l_i(t))}, I_i(0), B_i(t-), dN_i(t-)\\
\tag{P1}0<P(dN_i(t)=1 &|\mathbf{K_i(t)}, \mathbf{\mathcal{H}^o_i(t-)}), P(dN_i(t)=0 |\mathbf{K_i(t)}, \mathbf{\mathcal{H}^o_i(t-)}) <1 \\
\tag{P2}0<P(I_i(t)=1 |& \mathbf{K_i(t)}, \mathbf{\mathcal{H}^o_i(t-)}, dN_i(t)), P(I_i(t)=0 | \mathbf{K_i(t)}, \mathbf{\mathcal{H}^o_i(t-)}, dN_i(t))<1,
\end{align}
as well as no interference, and correct model specifications for the exposure, the outcome and the visit models, the proposed estimator resulting from the following estimating equation
\begin{align}
\mathbb{E} \left( \int_0^{\tau}\frac{ \mathbf{Y(t)} -  [\bm{\beta_s}' \mathbf{S(t)}]}{\mathbf{w(t|K(t))} \mathbf{sw_j(t| \mathcal{H}^o(t-))}  } \mathbf{dN(t)} \right) =\bm{0} \label{eq}
\end{align}
 is a two-step m-estimator. Two-step estimators often rely on substituting an estimate of a nuisance parameter in the estimating function for the parameter of interest (Newey and McFadden, 1994). One can use a first-step estimator for the nuisance parameters (e.g., here, the parameters from the IPT weights and from the monitoring weights). M-estimators of the parameters of interest are obtained by solving a sample average equation and often consist of the zero roots of an estimating equation.

For $\bm{\hat{\beta}_{TSE}}$ a two-step semiparametric estimator and $\bm{\beta_0}$ the vector of true parameters, Newey and McFadden (1994) show that $\sqrt{n}(\bm{\hat{\beta}_{TSE}-\beta_0)} \rightarrow N(\mathbf{0}, \mathbf{\Sigma})$, with
\begin{align}
\mathbf{\Sigma}=\mathbf{G_{\beta}^{-1}} \mathbb{E}\left[ \mathbf{ \left\{ g(o; \beta_0, \phi_0) - G_{\phi} M^{-1} m(o; \phi_0)\right\}}^{\otimes 2} \right]\mathbf{ G_{\beta}^{-1}}
\end{align}
where
\begin{align*}
\mathbf{G_{\beta}}=\mathbf{\mathbb{E}(\triangledown_{\beta} g(o; \beta_0, \phi_0))} \\
\mathbf{G_{\phi}}= \mathbf{\mathbb{E}(\triangledown_{\phi} g(o; \beta_0, \phi_0))} \\
\mathbf{M} =\mathbf{ \mathbb{E}(\triangledown_{\phi} m(o; \phi_0)) }
\end{align*}
for $\mathbf{o}$ the data, and $\mathbf{m(o; \phi_0)}$ and $\mathbf{g(o; \beta_0, \phi_0)}$ the estimating equations for the nuisance parameters $\bm{\phi}$ and the parameters of interest $\bm{\beta}$, respectively.\vspace{0.2cm}\\ In the case of interest, the nuisance parameters consist in the parameters from the intervention model and those from the intensity model. The estimating equations for the former are those corresponding to a logistic regression model for the exposure, and those for the latter correspond to equations from the Cox model for the visit intensity. 

\noindent We do not show how to compute the variance component due to the inverse monitoring weight but one could potentially develop that component of variance by using extensions of the Greenwood formula for the survivor function (Greenwood, 1926). In particular, the product integral used in our cumulated weight can be approximated by an exponential function since that we take the product of small quantities, in continuous. The delta method could be used in the development of the variance. However in practice, a nonparametric bootstrap (resampling on individuals) will provide a good estimate of the estimator variance. Alternatively, a ``robust'' variance may serve as a conservative estimate of the variance, similar to the case in marginal structural models (Hern\`an, Brumback and Robins, 2000).

\newpage

\textbf{Supplementary Material D}
\textbf{Details of the simulation studies} \vspace{0.2cm}\\

In the main study, we first simulated for each patient $i$ three baseline confounders $\left\{K_{1i},K_{2i},K_{3i}\right\}$ with $K_{1i}\sim \text{N}(1,1), K_{2i}\sim \text{Bernoulli}(0.55)$, and $K_{3i}\sim \text{N}(0,1)$. The intervention $I_i(t)$ was binary and time-fixed: $I_i\sim \text{Bernoulli}(p_{Ii})$ with $p_{Ii}=$ $\text{expit} \left( 0.5 + 0.8\hspace{0.02cm} K_{1i}+ 0.05\hspace{0.02cm} K_{2i} -1 \hspace{0.02cm}K_{3i}\right)$. One time-varying mediator $Z_i(\cdot)$ was generated, conditional on $I_i$. It was only updated whenever there was a new visit ($dN_i(\cdot)=1$), and was simulated as $Z_i(t)|I_i=1 \sim \text{N}(2,1)$ and $ Z_i(t)|I_i=0 \sim \text{N}(4,2^2)$
on those visit days. On other (non-visit) days, we denote the process by $Z_i(l_i(t))$, simply carrying forward the last observed value. Time was discretized over a grid of 0.01 units, from 0 to $\tau$. The intensity of monitoring at each time point over that grid was simulated as $\lambda_i(t| I_i,Z_i(l_i(t)))= 0.02 B_i(t) \exp \left( \gamma_1 I_i  + \gamma_2 Z_i(l_i(t)) \right)$. We used Bernoulli draws with probabilities proportional to these intensities to assign monitoring times (one draw per time point, for each time point over the grid). Whenever a new monitoring time occurred, the endogenous covariate process $\mathbf{Z(l(t))}$ was updated according to the simulation scheme given above (i.e.~depending on the value of the baseline intervention $I_i$), the outcome was simulated dependent on the gap time as detailed below, and then the gap time $B_i(t)$ was reset to 0. On each subsequent day when there was no visit, the gap time cumulated a value of 0.01 according to our discrete time grid.

We considered several different combinations of the parameters $(\gamma_1,\gamma_2)$ (see Table 1), so as to vary the strength of the selection bias due to the visit process. The outcome $Y_i(t)$ was generated according to $Y_i(t)= 0.2 B_i(t) + 1\hspace{0.05cm} I_i - 0.8\hspace{0.05cm} \left(Z_i(l_i(t)) - E\left[Z_i(l_i(t)) |I_i\right]\right) +  0.4\hspace{0.05cm} K_{1i}+ 0.05\hspace{0.05cm}K_{2i} -0.6 \hspace{0.05cm} K_{3i} +  \epsilon_i(t)$ with $ \epsilon_i(t) \sim \text{N}(0, 0.5^2)$. The re-centering of $Z_i(l_i(t))$ in the outcome model ensures that we estimate the target marginal intervention effect. Monitoring times were drawn up until the maximum follow-up time $\tau$, which we fixed to $\tau=5$. Data were simulated to correspond to a study cohort of 500 patients. For each patient, the follow-up time was ``censored" (stopped) at time $C_i$, with $C_i\sim \text{Uniform}(\tau/2, \tau)$; the censoring was non-informative. A total of $1000$ replicate datasets were simulated for each simulation study scenario.

\newpage

\textbf{Supplementary Material E}
\textbf{Results of the main simulation study, including the average number of visits and estimated parameters in the visit model} \vspace{0.2cm}\\

\begin{table}[H]
 \begin{center}
\caption{Main analysis: Estimated parameters, average number of visits and mean absolute bias for the estimators compared for 1000 simulations with $\tau=5$, $n=500$.}
\begin{tabular}{cccccccc c }
 \hline \hline
  $\bm{\gamma}$   &$\hat{\bm{\gamma}}$ &$\overline{N}(\tau)$    &\multicolumn{6}{c} {Mean absolute bias of the estimator}   \\
      &   &   $I=0,1$ & $\hat{\beta}_{LS}$&$\hat{\beta}_{IPT}$&$\hat{\beta}_{IH}$&$\hat{\beta}_{USW}$  & $\hat{\beta}_{SW1}$   & $\hat{\beta}_{SW2}$\\ \hline 
-0.3; 0.1&-0.3; 0.1& 1.9, 2.9 &0.35&0.37&0.12&0.04&0.14&0.03\\
-0.2; 0.2&-0.2; 0.2&2.5, 3.5  &0.49 &0.24&0.11&0.00&0.09&0.01\\
-0.1; 0.2&-0.1; 0.2&3.0, 3.9  &0.64 &0.08&0.09&0.14&0.00&0.02 \\
-0.1; -0.3&-0.1; -0.3&3.1, 2.9 & 0.69&0.01&0.11&0.07&0.01&0.03 \\
0; 0&0; 0&3.9, 3.9  &0.73 &0.01&0.03&0.16&0.01&0.01\\
0.1; -0.3&0.1; -0.3&4.8, 3.7 &0.69 &0.03&0.11&0.19&0.04&0.02 \\
0.2; -0.2&0.2; -0.2&6.0, 4.3  &0.64 &0.12&0.26&0.18&0.19&0.02 \\
0.3; 0.2& 0.3; 0.2& 7.1, 5.9& 0.67&0.08 &0.34 &0.24&0.30&0.05 \\   \hline
 \end{tabular}
 \label{tabtest}
\end{center}
\end{table}

\newpage

\textbf{Supplementary Material F}
\textbf{Results of all sensitivity analyses} \vspace{0.2cm}\\

\begin{table}[H]
\caption{Simulation study results for sensitivity analysis 1 with $\tau=5$, $n=500$ patients, 1000 simulations, $\alpha_i(t) = 0.2\hspace{0.05cm} B_i(t)$. A constant intercept was fitted in the outcome model, rather than a cubic spline as a function of gap time }
\begin{tabular}{cccccccccccccc }
 \hline \hline
  ${\bm{\gamma}}^\dagger$ &$\overline{N(\tau)}$    &\multicolumn{6}{c} {Mean absolute bias $\hat{\beta}$} &\multicolumn{6}{c} {Empirical variance $\hat{\beta}$} \\    
       &   $I=0,1$ & $\hat{\beta}_{LS}$&$\hat{\beta}_{IPT}$&$\hat{\beta}_{IH}$&$\hat{\beta}_{USW}$  & $\hat{\beta}_{SW1}$  & $\hat{\beta}_{SW2}$& $\hat{\beta}_{LS}$&$\hat{\beta}_{IPT}$&$\hat{\beta}_{IH}$&$\hat{\beta}_{USW}$& $\hat{\beta}_{SW1}$& $\hat{\beta}_{SW2}$\\ \hline
 a&  1.9, 2.9& 0.45&0.25&0.18&0.08&0.16& 0.04  & 0.03&0.08 &0.17 &1.56 &0.13 &0.09 \\  
b&3.0, 3.9& 0.65 &0.05&0.11&0.09&0.01& 0.01&0.02&0.05 & 0.10&1.16 &0.06 & 0.05  \\  
c&3.9, 3.9 & 0.72 &0.03&0.04&0.14&0.03& 0.03& 0.02 &0.06 &0.10 &0.59 &0.06 &0.06 \\  
d& 4.8, 3.7 & 0.79 &0.08&0.06&0.17&0.02&0.02 & 0.02 &0.04 &0.07&0.42 &0.05 &0.05\\  
e&7.1, 5.9& 0.79&0.08&0.31&0.25&0.25& 0.05&0.02 &0.03 &0.06 &0.29 &0.08 &0.06 \\  
 \hline
\end{tabular}
 \label{app4:res0}
 
 \scriptsize{$\dagger$: a. (-0.3, 0.1); b. (-0.1, 0.2); c. (0, 0); d. (0.1, -0.3); e. (0.3, 0.2). }
 
 \end{table}

\begin{table}[H]
 
\caption{Simulation study results for sensitivity analysis 2 with $\tau=10$, $n=500$ patients, 1000 simulations, $\alpha_i(t) = 0.2\hspace{0.05cm} B_i(t)$ }
\begin{tabular}{cccccccccccccc }
 \hline \hline
  ${\bm{\gamma}}^\dagger$ &$\overline{N(\tau)}$    &\multicolumn{6}{c} {Mean absolute bias $\hat{\beta}$} &\multicolumn{6}{c} {Empirical variance $\hat{\beta}$} \\    
       &   $I=0,1$ & $\hat{\beta}_{LS}$&$\hat{\beta}_{IPT}$&$\hat{\beta}_{IH}$&$\hat{\beta}_{USW}$  & $\hat{\beta}_{SW1}$  & $\hat{\beta}_{SW2}$& $\hat{\beta}_{LS}$&$\hat{\beta}_{IPT}$&$\hat{\beta}_{IH}$&$\hat{\beta}_{USW}$& $\hat{\beta}_{SW1}$& $\hat{\beta}_{SW2}$\\ \hline
 a&  4.2, 6.2&0.39&0.35&0.14&0.02&0.15&0.02&0.01&0.03&0.06&1.40&0.08&0.03\\  
b&6.6, 8.1& 0.65 &0.07&0.10&0.15&0.01&0.01&0.01&0.02&0.04&0.67&0.03&0.02\\  
c&8.1, 8.1 &0.72&0.02&0.02&0.13&0.02&0.02&0.01&0.02&0.03&0.25&0.02&0.02\\  
d& 10.0, 7.7 &0.68&0.05&0.12&0.12&0.04&0.02&0.01&0.02&0.03&0.17&0.02&0.02\\  
e& 13.5, 11.2&0.64&0.11&0.38&0.06&0.30&0.05&0.01&0.02&0.03&0.15&0.04&0.04\\  
 \hline
\end{tabular}
 \label{app4:res1}
 
 \scriptsize{$\dagger$: a. (-0.3, 0.1); b. (-0.1, 0.2); c. (0, 0); d. (0.1, -0.3); e. (0.3, 0.2). }

\end{table}

\begin{table}[H]
 
\caption{Simulation study results for sensitivity analysis 3 with $\tau=5$, $n=500$ patients, 1000 simulations, $\alpha_i(t) = 0.2\hspace{0.05cm} B_i(t)$, and with the process $Z(\cdot)$ depending on the cumulative number of previous visits }
\begin{tabular}{cccccccccccccc }
 \hline \hline
  ${\bm{\gamma}}^\dagger$ &$\overline{N(\tau)}$    &\multicolumn{6}{c} {Mean absolute bias $\hat{\beta}$} &\multicolumn{6}{c} {Empirical variance $\hat{\beta}$} \\    
       &   $I=0,1$ & $\hat{\beta}_{LS}$&$\hat{\beta}_{IPT}$&$\hat{\beta}_{IH}$&$\hat{\beta}_{USW}$  & $\hat{\beta}_{SW1}$  & $\hat{\beta}_{SW2}$& $\hat{\beta}_{LS}$&$\hat{\beta}_{IPT}$&$\hat{\beta}_{IH}$&$\hat{\beta}_{USW}$& $\hat{\beta}_{SW1}$& $\hat{\beta}_{SW2}$\\ \hline
 a&  1.9, 2.8&0.35&0.38&0.12&0.05&0.16&0.04&0.02&0.05&0.10&1.18&0.09&0.05\\  
 b&3.1, 3.8  &0.63&0.09&0.09&0.09&0.02&0.00&0.01&0.04&0.06&0.74&0.04&0.04\\  
c& 3.9, 3.9 &0.72&0.02&0.04&0.13&0.02&0.02&0.01&0.02&0.05&0.33&0.02&0.02\\  
d&4.9, 3.7  &0.69&0.04&0.11&0.19&0.05&0.01&0.01&0.02&0.04&0.24&0.02&0.03\\  
e& 8.1, 6.6 &0.65&0.10&0.36&0.18&0.32&0.05&0.01&0.02&0.04&0.19&0.04&0.04\\  
 \hline
\end{tabular}
 \label{app4:res2}

  \scriptsize{$\dagger$: a. (-0.3, 0.1); b. (-0.1, 0.2); c. (0, 0); d. (0.1, -0.3); e. (0.3, 0.2). }
\end{table}

\begin{table}[H]
 
\caption{Simulation study results for sensitivity analysis 4 with $\tau=5$, $n=500$ patients, 1000 simulations, $\alpha(t)= \alpha = 0.02.$}
\begin{tabular}{cccccccccccccc }
 \hline \hline
  ${\bm{\gamma}}^\dagger$ &$\overline{N(\tau)}$    &\multicolumn{6}{c} {Mean absolute bias $\hat{\beta}$} &\multicolumn{6}{c} {Empirical variance $\hat{\beta}$} \\    
       &   $I=0,1$ & $\hat{\beta}_{LS}$&$\hat{\beta}_{IPT}$&$\hat{\beta}_{IH}$&$\hat{\beta}_{USW}$  & $\hat{\beta}_{SW1}$  & $\hat{\beta}_{SW2}$& $\hat{\beta}_{LS}$&$\hat{\beta}_{IPT}$&$\hat{\beta}_{IH}$&$\hat{\beta}_{USW}$& $\hat{\beta}_{SW1}$& $\hat{\beta}_{SW2}$\\ \hline
 a& 1.9, 2.9 &0.36&0.34&0.17&0.02&0.10&0.00&0.02&0.04&0.07&1.35&0.06&0.04\\  
b & 3.1, 3.9 &0.63&0.09&0.09&0.10&0.01&0.01&0.01&0.04&0.07&0.77&0.04&0.04\\  
 c&3.9, 3.9  &0.73&0.02&0.03&0.13&0.02&0.02&0.01&0.03&0.04&0.36&0.03&0.03\\  
d& 4.8, 3.7 &0.70&0.03&0.11&0.18&0.04&0.01&0.01&0.03&0.04&0.26&0.03&0.03\\  
e& 7.1, 5.9&0.66&0.09&0.34&0.22&0.30&0.06&0.01&0.02&0.04&0.20&0.04&0.04\\  
 \hline
\end{tabular}
 \label{app4:res3}
 
  \scriptsize{$\dagger$: a. (-0.3, 0.1); b. (-0.1, 0.2); c. (0, 0); d. (0.1, -0.3); e. (0.3, 0.2). }
 
\end{table}

\newpage
\textbf{Supplementary Material G}
\textbf{Comparison of the bootstrap and the empirical variance of the estimators} \vspace{0.2cm}\\

\begin{table}[H]
 \caption{ Comparison of bootstrap and empirical variance for all simulation studies in the main analysis (studies with $\tau=5$, $n=500$, 1000 simulations) }
\begin{tabular}{ cccccccccccccc }
 \hline \hline
Intercept &  $\bm{\gamma}$   & \multicolumn{6}{c} {Empirical variance of $\hat{\beta}$} &\multicolumn{6}{c} {Bootstrap variance $\hat{\beta}$} \\    
fitted &    no.$^\dagger$ & $\hat{\beta}_{LS}$&$\hat{\beta}_{IPT}$&$\hat{\beta}_{IH}$&$\hat{\beta}_{USW}$  & $\hat{\beta}_{SW1}$& $\hat{\beta}_{SW2}$ & $\hat{\beta}_{LS}$&$\hat{\beta}_{IPT}$&$\hat{\beta}_{IH}$&$\hat{\beta}_{USW}$& $\hat{\beta}_{SW1}$& $\hat{\beta}_{SW2}$\\ \hline
 
Constant&1&0.01&0.04&0.09&1.25&0.07&0.05&0.03&0.08&0.17&1.56&0.13&0.09\\  
&2&0.01&0.04&0.08&0.98&0.06&0.05&0.03&0.06&0.11&1.19&0.09&0.07\\ 
&3&0.01&0.03&0.06&0.81&0.03&0.03&0.02&0.05&0.10&1.16&0.06&0.05\\ 
&4&0.01&0.03&0.06&0.43&0.04&0.03&0.02&0.06&0.10&0.63&0.07&0.06\\ 
&5&0.01&0.03&0.05&0.38&0.03&0.03&0.02&0.06&0.10&0.59&0.06&0.06\\ 
&6&0.01&0.02&0.05&0.25&0.03&0.03&0.02&0.04&0.07&0.42&0.05&0.05\\ 
&7&0.01&0.02&0.05&0.23&0.04&0.03&0.02&0.03&0.07&0.38&0.06&0.05\\ 
&8& 0.01&0.02&0.05&0.20&0.05& 0.04& 0.02&0.03 &0.06&0.29&0.08&0.06 \\
 \hline
Cubic  &1&0.02&0.04&0.10&1.21&0.07&0.04&0.03&0.07&0.16&1.56&0.13&0.08\\  
spline&2&0.01&0.04&0.08&1.02&0.05&0.04&0.03&0.06&0.11&1.23&0.08&0.06\\ 
&3&0.01&0.03&0.06&0.72&0.03&0.03&0.02&0.05&0.09&1.11&0.06&0.05\\ 
&4&0.01&0.03&0.06&0.39&0.04&0.03&0.02&0.05&0.08&0.59&0.06&0.05\\ 
&5&0.01&0.03&0.06&0.35&0.03&0.03&0.02&0.05&0.09&0.56&0.05&0.05\\ 
&6&0.01&0.03&0.04&0.24&0.03&0.03&0.02&0.04&0.06&0.40&0.05&0.05\\
&7&0.01&0.03&0.05&0.24&0.03&0.04&0.02&0.03&0.06&0.36&0.05&0.05\\ 
&8& 0.01&0.02&0.04 & 0.18&0.05 &0.04& 0.02&0.03&0.06& 0.26& 0.07&0.05 \\
 \hline
\end{tabular}
 \label{tab:res1}

$\dagger$. \scriptsize{\textbf{1.} (-0.3, 0.1); \textbf{2.} (-0.2, 0.2); \textbf{3.} (-0.1, 0.2); \textbf{4.} (-0.1, -0.3); \textbf{5.} (0, 0); \textbf{6.} (0.1, -0.3); \textbf{7.} (0.2, -0.2);  \textbf{8.} (0.3, 0.2).}
 
\end{table}
\newpage
\textbf{Supplementary Material H}
\textbf{Table of baseline characteristics stratified by intervention group, in the analysis of CPRD data }\vspace{0.2cm}\\

 \begin{table}[H]
 \begin{center}
\caption{Baseline characteristics of the study cohort stratified by treatment at cohort entry (n=246,503), Clinical Practice Research Datalink, United Kingdom, 1998-2017 }
\begin{tabular}{lcc }
 \hline \hline
   &\multicolumn{2}{c}{Treatment} \\
Variable$^1$& Citalopram& Fluoxetine \\ \hline
 
  BMI, mean &26.8& 26.8\\
Age, mean &43.4& 40.7 \\
 Female sex &63.6 & 66.1 \\
  IMD quintile$^2$, mean &3.01&3.08 \\
 Smoking status & & \\
$\hspace{0.5cm}$ Ever& 50.8& 48.7\\
$\hspace{0.5cm}$ Never&34.0 &31.3 \\
$\hspace{0.5cm}$ Unknown & 15.2&20.1 \\
 Diabetes &5.3 &4.3\\
Alcohol abuse &8.0& 6.6 \\
Anxiety or GAD &30.5&22.1 \\
Other psychiatric diseases & & \\
$\hspace{0.5cm}$ Schizophrenia & 1.4 & 1.1 \\
$\hspace{0.5cm}$ Bipolar disorder & 0.8 & 0.7 \\
$\hspace{0.5cm}$ Autism spectrum disorder & 0.2 &0.1 \\
$\hspace{0.5cm}$ Obsessive compulsive disorder & 0.5 & 0.6 \\
  Antipsychotic drugs & 11.7 & 10.6 \\
  Benzodiazepine drugs &19.7&17.1 \\
  Lipid lowering drugs &7.6&5.0 \\
 \hline

\end{tabular}
 \label{tab:baseline}

 \scriptsize{Abbreviations: BMI, Body mass index; IMD, Index of multiple deprivation; GAD, Generalized anxiety disorder.\\
  1. In \% unless otherwise stated.\\
  2. The IMD was available in the format of quintiles, with the greater quintile being the most deprived}
 \end{center}
 \end{table}
\newpage
\textbf{Supplementary Material I}

\textbf{Multivariate outcome model in the analysis of the CPRD data }\vspace{0.2cm}\\

\noindent We present below the coefficients for each covariate in a linear multivariate outcome model for the outcome BMI (no inverse weight is incorporated in this model).
  \begin{table}[H]
 \begin{center}
\caption{Coefficients for each covariate included in the conditional outcome mean model (except the intercept and spline on time), Clinical Practice Research Datalink, United Kingdom, 1998-2017 }
\begin{tabular}{ lcccc }
 \hline \hline
Variable &Coefficient& Robust 95\% CI\\ \hline
 
    Citalopram (Ref.: Fluoxetine) & -0.55&-0.67, -0.43* \\
    Age at baseline       &-0.02&-0.03, -0.02*  \\
     Sex (Ref.: Female)     & -0.58 & -0.70, -0.46* \\
  IMD at baseline        &0.40&0.35, 0.44* \\
  Smoking (Ref.: Never)  && \\
$\hspace{0.5cm}$  Ever    &-0.53& -0.68, -0.38*\\
$\hspace{0.5cm}$ Missing      &0.03& -0.12, 0.18\\
 Diabetes     &4.21& 4.03, 4.39*\\
  Alcohol abuse  & -0.66& -1.13, -0.20*\\
     Anxiety or GAD &-0.22&-0.43, -0.01* \\
     Psychiatric diagnosis    &0.24&-0.49, 0.96\\
   Number of hospitalisations in prior month         &-0.69&-0.86, -0.52* \\
  Antipsychotic drugs         &-0.05& -0.43, 0.33\\
   Benzodiazepine drugs         &-1.55&-1.77, -1.33* \\
  Lipid lowering drugs         &1.66&1.48, 1.85* \\
 \hline
\end{tabular}
 \label{tab:res3b}

  \scriptsize{  Abbreviations: IMD, Index of Multiple Deprivation; GAD, Generalized Anxiety Disorder.\\
  * Confidence interval does not contain 0.}
 \end{center}
 \end{table}
  
\end{document}